\documentclass[showpacs,10pt,aps,prd,reprint,onecolumn,notitlepage,longbibliography]{revtex4-1}
\usepackage{amsmath}
\usepackage{bbm}
\usepackage{amsfonts}
\usepackage{amssymb}
\usepackage{bm}
\usepackage{tensor}
\usepackage[all,cmtip]{xy}
\usepackage{accents}
\usepackage[bookmarks=false]{hyperref}

\providecommand{\ed}{\mathrm{d}}

\providecommand{\mf}{\bm{F}}

\providecommand{\mJ}{\bm{J}}

\providecommand{\mc}{\bm{\chi}}

\begin{document}

\title{Covariant electrodynamics in linear media: Optical metric}
\author{Robert T. Thompson}
\email{robert@cosmos.phy.tufts.edu}
\affiliation{Institute of Applied Physics, 
Karlsruhe Institute of Technology (KIT), 76131 Karlsruhe, Germany} 

\begin{abstract}
While the postulate of covariance of Maxwell's equations for all inertial observers led Einstein to special relativity, it was the further demand of general covariance -- 
form invariance under general coordinate transformations, including between accelerating frames -- that led to general relativity.
Several lines of inquiry over the past two decades, notably the development of metamaterial-based transformation optics, has spurred a greater interest in the role of geometry and space-time covariance for electrodynamics in ponderable media.
I develop a generally covariant, coordinate-free framework for electrodynamics in general dielectric media residing in curved background space-times. 
In particular, I derive a relation for the spatial medium parameters measured by an arbitrary timelike observer.  In terms of those medium parameters I derive an explicit expression for the pseudo-Finslerian optical metric of birefringent media and show how it reduces to a pseudo-Riemannian optical metric for nonbirefringent media.
This formulation provides a basis for a unified approach to ray and congruence tracing through media in curved space-times that may smoothly vary among positively refracting, negatively refracting, and vacuum.
\end{abstract}

\pacs{}
\maketitle

\section{Introduction}
The standard vectorial representation of Maxwell's equations 
\begin{subequations} \label{Eq:Maxwell3D}
	\begin{align} 
	\nabla\cdot\vec{B} & = 0, & \nabla\times\vec{E} +\frac{\partial\vec{B}}{\partial t} & = 0,  \label{Eq:MaxwellHomog} \\
	\nabla\cdot\vec{D} & = \rho, & \nabla\times\vec{H} - \frac{\partial\vec{D}}{\partial t} & = \vec{j}, \label{Eq:MaxwellInhomog}
	\end{align}
\end{subequations}
has been the electrodynamics workhorse since it was codified by Heaviside well over a century ago. 
But this 3-vector representation harbors a hidden assumption: that space-time is Minkowskian.
We may forgive Heaviside this trespass, coming 30 years before Einstein's seminal paper on general relativity, but
no description of electrodynamics would be truly complete without incorporating the properties of the underlying space-time in which the fields exist.

While the relativistic nature of electrodynamics is fully acknowledged, the general space-time formulation is rarely employed in applications.  
Standard, widely regarded texts on electrodynamics or relativity typically only discuss tensorial electrodynamics of the vacuum \cite{Jackson,MTW,Baez}.
But ponderable media do not displace space-time, as a whale displaces water. Rather, they are completely surrounded and permeated by the geometry of space-time -- even the whale in water feels gravity.
Thus no theory of macroscopic electrodynamics within media is truly complete without incorporating the properties of the underlying space-time in which the medium resides.

Here on Earth, most practical applications of electrodynamics do not require the heavy artillery of general relativity, but there are still good reasons to be interested in a space-time covariant formulation of electrodynamics within media.
In particular, it is closely related to a convergence of ideas in transformation optics, analog gravity, premetric electrodynamics, and Lorentz violating space-times that have emerged over the past 15 years.

Transformation optics uses ideas about the similarity of the refractive properties of dielectric media with the light-bending properties of curved space-times to design optical media with unusual properties or functionality, such as negatively refracting media and invisibility cloaks \cite{Pendry2000prl,Pendry2006sc,Leonhardt2006sc,Leonhardt2006njp1,McCall2011jo}.
The advent of structured composite metamaterials that possess some of the unnatural properties required to actually realize these unusual devices \cite{Smith2000prl,Smith2000prl1,Schurig2006sc} provides an exciting demonstration
that the geometrical, space-time manifold aspects of electrodynamics and light propagation are not just an academic abstraction, but can have real applications in engineering.
Progress understanding and developing this new theoretical technology relies on the type of covariant formulation of electrodynamics in media studied here \cite{Teixeira2007motl,Bergamin2008pra,Leonhardt2009po,Bergamin2011prb,Thompson2011jo1,Thompson2011jo2,Paul2012oe,Thompson2015pra,Gratus2016njp}.

Analog models of curved space-times seek to replicate certain aspects of extreme gravitational systems, such as light propagation near black holes, in a nongravitational laboratory-accessible system, such as sound or surface waves in flowing fluids \cite{Unruh1981prl,Schutzhold2002prd,Weinfurtner2011prl,Barcelo2005lrr}, or, more relevantly, light propagating through a suitable dielectric medium \cite{DeFelice1971gerg,Reznik2000prd,Schutzhold2002prl,Greenleaf2007prl,Philbin2008sc,Narimanov2009apl,Thompson2010prd,Smolyaninov2011jo,Bini2013gerg,FernandezNunez2016pla}.
Once again it is the space-time covariant formulation of electrodynamics in media and the geometrical aspects of light propagation on manifolds that enables detailed study of dielectric analog space-times, and their limitations \cite{Fathi2016prd}.

Premetric electrodynamics is based on the observation that the space-time metric does not enter Maxwell's equations directly, but only through its associated Hodge dual \cite{Hehl2002ijmpa,Hehl}.  By promoting the Hodge dual in Maxwell's equations to an independent structural field on a metric-free manifold, one may gain new insights into the structure of electrodynamics.

Lastly, Lorentz-violating space-times play a significant role in theories of quantum gravity and physics beyond the standard model, and it has been shown that Lorentz violation in effective field theories is connected to pseudo-Finsler geometries \cite{Kostelecky1989prd,Girelli2007prd,Kostelecky2011plb}.
The Lorentzian nature of space-time was inspired by Maxwell's equations in vacuum, but it is known that the birefringence exhibited by general linear media is connected to pseudo-Finsler geometries \cite{Perlick_Ray_Optics,Lammerzahl2004prd,Skakala2009jp}, providing a natural setting for further musings on the nature of space-time.

In short, Finsler geometry is just Riemannian geometry without the quadratic restriction \cite{Chern1996ams}.  
Riemannian metric geometry is based on a line element of the form
\begin{equation}
 ds = F(x^1,\dots,x^n; dx^1,\dots,dx^n),
\end{equation} 
where $F$ is a function on the tangent bundle such that
\begin{equation}
 F^2 = g_{\alpha\beta}(x)dx^{\alpha}dx^{\beta},
\end{equation}
which provides the metric structure
\begin{equation}
 g_{\alpha\beta}(x) = \frac{\partial F^2}{\partial x^{\alpha} \partial x^{\beta}}.
\end{equation}
Finsler geometry relaxes this quadratic restriction on $F$, so that one may still define the metric in a similar way, but it is no longer independent of the cotangent basis.  
In other words, the metric depends not only on the point on the manifold, but also on which direction you look \cite{Dahl2006thesis}.

\subsection{Goal of this paper}

The goal of this paper is to construct a covariant form of electrodynamics in linear media residing in a curved background space-time.  
Some elements of this formalism have been introduced and used in several papers studying transformation optics and dielectric analog space-times \cite{Thompson2010prd,Thompson2011jo1,Thompson2017prd}.
Here, those initial elements are extended, given a rigorous derivation, and combined into a comprehensive framework.
In particular, the relationship between the four-dimensional $\binom{2}{2}$ material tensor $\bm{\chi}$ and the usual spatial material parameters $\bm{\varepsilon}$, $\bm{\mu}$, $\tensor[^h]{\bm{\gamma}}{}$ and $\tensor[^e]{\bm{\gamma}}{}$ is derived \textit{ab initio}, and an expression for the pseudo-Finslerian optical metric is derived for quite general media and given in terms of the aforementioned usual spatial material parameters.

In keeping with the tenets of general covariance, the sought after formalism should be independent of the chosen coordinates in the sense that there exists a well-defined method of shifting between coordinates, and therefore should be expressible entirely in terms of four-dimensional tensorial objects and the metric.
The exterior calculus of differential forms is particularly well suited to the study of electrodynamics and will be adopted here.  

Certainly this is not the first paper to study the covariant form of electrodynamics in media, which has a long history starting with Minkowski, Gordon, and Tamm in the early part of the last century \cite{Minkowski1910ma, Gordon1923,Tamm1924}, followed by Balazs, Quan, Plebanski, Post, and Ehlers \cite{Balazs1955josa,Quan1962ltrg,Plebanski1960pr,Post,Ehlers1966pgr} in the middle of the last century, and in more modern times by Perlick \cite{Perlick_Ray_Optics}, Novello and Salim \cite{Novello2001prd}, Novello, Lorenci, Salim, and Klippert \cite{Novello2000prd}, Novello and Bittencourt \cite{Novello2012prd}, Visser, Barcel\'{o}, and Liberati \cite{Visser2003Novello}, Balakin and Zimdahl \cite{Balakin2005gerg}, and notably as part of the premetric community by Hehl, Obukhov, Rubilar, L\"{a}mmerzahl, and Itin \cite{Hehl2002ijmpa,Lammerzahl2004prd,Itin2009jpa}.
Indeed, after more than 100 years it is still a fruitful area of research, with a very recent and thorough analysis by Schuster and Visser that is strongly related to the work pursued here \cite{Schuster2017prd}, where they also consider a background metric and introduce the space-time splitting with respect to a timelike observer.

There are a number of mathematical similarities and common themes between these previous works and what is presented here.  
For example, differential forms are also used extensively by the premetric community, so some of the equations here have close counterparts in premetric electrodynamics.

It is quite common in the literature to adopt a 6-vector approach when dealing with electrodynamics in media; in other words, instead of the usual space-time formulation whereby $(\vec{D},\vec{H})$ and $(\vec{E},\vec{B})$ are select components of two-forms, they are put into two six-component vectors.
While there is nothing invalid about such an approach, it does suffer from two drawbacks: 1) it subtly assumes an explicit 3+1 decomposition of the space-time, which means that the field components have been rather rigidly defined, and makes the transformation properties somewhat less obvious, and 2) it makes the theory less compatible with other generally relativistic calculations one might want to perform, such as examining the Raychaudhuri equation for a congruence of light passing through ponderable media \cite{Thompson2017prd}.
Instead, everything done here is fully covariant, four-dimensional, coordinate-free, and largely index-free.  
Of course, reference to the usual spatial parameters naturally requires a 3+1 split, but here this is done with respect to a timelike vector field that need not align in any special way with a choice of coordinates.
I believe this point of view provides additional insight into the structure of the theory by highlighting the role of the observer who must ultimately be making measurements on the fields and the medium, and is a mechanism also employed in Refs.~\cite{Balakin2005gerg} and \cite{Schuster2017prd}.

The paper is organized as follows. Section \ref{Sec:3VectorEM} briefly reviews the crucial aspects of 3-vector electrodynamics in dielectric media that must be generalized to the four-dimensional setting.  Section \ref{Sec:Notation} introduces the mathematical notation that will be used throughout the rest of the paper.  Section \ref{Sec:Electrodynamics} introduces the tensorial version of electrodynamics in media and the differential and integral forms of Maxwell's equations, while Sec.~\ref{Sec:4Dto3D} derives the relationship between the four-dimensional tensors and the ``usual'' spatial, or transverse, material tensors.  Section \ref{Sec:GeometricOptics} presents the geometric optics limit of the theory, while Sec.~\ref{Sec:OpticalMetric} derives the pseudo-Finslerian optical metric in this limit, and Sec.~\ref{Sec:RiemannReduction} shows how the pseudo-Finslerian optical metric becomes pseudo-Riemannian in certain circumstances.  Finally, Sec.~\ref{Sec:Kinematics} describes how all of these derivations relate to something tangible like ray tracing.  I conclude with Sec.~\ref{Sec:Conclusions}.

\section{3-vector classical electrodynamics in media} \label{Sec:3VectorEM}
Since the goal of this paper is to study the covariant formulation of electrodynamics in dielectric media, it is useful to first review the familiar aspects of 3-vector electrodynamics in media that we wish to treat.
In the three-dimensional Cartesian coordinates of flat Minkowski space-time, Maxwell's equations take the form of Eqs.~(\ref{Eq:Maxwell3D})
where $\vec{E}$ and $\vec{D}$ are, respectively, the electric field and electric flux density, and $\vec{B}$ and $\vec{H}$ are, respectively, the magnetic flux density and the magnetic field, and where  the speed of light has been scaled to $c=1$.
The electric and magnetic fields $\vec{E}$ and $\vec{B}$ may be derived from scalar and vector potentials $\phi$ and $\vec{A}$ by
\begin{align} \label{Eq:3DPotentials}
 \vec{B} & = \nabla\times\vec{A}, & \vec{E} & = -\nabla \phi - \frac{\partial\vec{A}}{\partial t}.
\end{align}
However, as far as finding solutions is concerned, the set of Eqs.~(\ref{Eq:Maxwell3D}) and (\ref{Eq:3DPotentials}) is incomplete.  A supplemental condition is required on the set of fields $\{\vec{E},\vec{B},\vec{D},\vec{H}\}$,   
which is typically provided through the basic constitutive relations
\begin{equation} \label{Eq:3VecConstitutives}
 \vec{D} = \varepsilon \vec{E}, \quad \vec{B} = \mu \vec{H}.
\end{equation}

In vacuum, $\varepsilon=\varepsilon_0$ and $\mu=\mu_0$ are constants (and for $c=1$ we have $\varepsilon_0=\mu_0=1$), but inside dielectric media the permittivity $\varepsilon$ and permeability $\mu$ may not only be inhomogeneous but may not even be isotropic, described instead by matrix-valued functions.
The reduction of dielectric media to a set of supplemental parameters in Maxwell's equations belies an underlying quantum field theoretic, or at least microscopic, description of the interaction of electromagnetic fields and matter \cite{Loudon,deGroot1965phys}, but at the macroscopic level it is sufficient to use an effective theory that accounts for the average atomic response to the applied fields.
The basic idea of this model is that an electron in an atom is slightly displaced by an applied electric field.  
The electron feels a restoring force determined by the nucleus.  
The exact form of the restoring force may be modulated by the presence of other electrons and neighboring dipole moments, but in the ``Lorentz'' approximation it is modeled as being of Hooke's law variety -- linear in the displacement.  
Thus the applied electric field induces a dipole moment $\vec{p}$, the magnitude of which depends, to lowest order, linearly on both the applied field and the effective spring constant.  
Sufficiently far from the dipole moment the dipole field is proportional to $\vec{p}$, and the net electric field is
\begin{equation} \label{Eq:ENet}
 \vec{E}_{\text{net}}=\vec{E}_{\text{applied}}+\vec{p} = \vec{E}_{\text{applied}}+\chi_{E}\vec{E}_{\text{applied}} = (1+\chi_{E})E_{\text{applied}}
\end{equation}
where the electric susceptibility $\chi_E$ is linear in the average effective spring constant for the material.  
This may readily be extended to anisotropic and inhomogeneous material responses by changing the scalar $\chi_E$ into a position dependent matrix $\bar{\bar{\chi}}_E(\vec{x})$.  
The matrix nature of $\bar{\bar{\chi}}_E$ means the restoring force felt by the electron depends on the direction of displacement, and the value of $\bar{\bar{\chi}}_E(\vec{x})$ represents an averaging  over a small region around the point $\vec{x}$.

In the aggregate theory the electric flux density is related to the electric field via a constitutive relation that is typically written as
\begin{equation} \label{Eq:DEPconst}
 \vec{D} = \vec{E} + \vec{P}
\end{equation}
where the polarization density $\vec{P}$ is found through a suitable averaging procedure of the dipole moment density.
So far we have only considered $\vec{P}$ as linear in $\vec{E}$ for the simple Hooke's law model.  
More generally, one could also consider anharmonic dipole behavior and higher order multipole moment contributions to $\vec{D}$ rather than just linearized dipoles, and 
each multipole contribution to the total polarization can be expanded in terms of $\vec{E}$ so that \cite{Condon1932josa,Satten1957jcp,deLange2003prsla}
\begin{equation} \label{Eq:PolarizationVec}
 D^i = \chi\indices{^i_j}E^j + \gamma\indices{^{ij}_k}\nabla_jE^k + \zeta\indices{^i_{jk}}E^jE^k + \cdots
\end{equation}
Linear media for which $\bm{\gamma}$ is identically zero are said to be \textit{linearly anisotropic}.  
In linearly anisotropic media the eigenstates of propagation are states of linear polarization. 
Linear media for which $\bm{\chi}= 0$ and $\bm{\gamma}\neq 0$ are said to be \textit{circularly anisotropic}.  
In circularly anisotropic media the eigenstates of propagation are states of circular polarization.  
In a general medium with $\bm{\chi}\neq 0$ and $\bm{\gamma}\neq 0$ the eigenstates of propagation are states of elliptical polarization.

Similar to the electric response, ponderable media may exhibit a magnetic response to an applied field, which in the aggregate is usually expressed through the corresponding relation
\begin{equation} \label{Eq:BHMconst}
 \vec{B} = \vec{H} + \vec{M}
\end{equation}
where $\vec{M}$ is the magnetization vector.

\section{Some remarks on mathematical notation} \label{Sec:Notation}
The next step is to generalize the Heavisidian 3-vector mathematics of classical electrodynamics to the tensorial quantities that have meaning in a four-dimensional space-time.
As much as possible, I will endeavor to use an index-free notation to eliminate unnecessary clutter and to ensure that the methodology is truly coordinate independent.
The purpose of this section is to clarify the notation that is used throughout the following sections, and in particular to help make the notation accessible to those who have experience with electrodynamics from the 3-vector perspective but limited experience with general relativity.  
There are many excellent resources providing more detailed explanation of the mathematical concepts being used, such as Refs.~\cite{Tu_Diff_Geom,Tu_Intro_Manifolds}. 
A concise introduction to the required topics of differential geometry used for electrodynamics, with a focus on transformation optics, is available in Ref.~\cite{Thompson2012aiep} or Ref.~\cite{Leonhardt2009po}.

In 3-vector notation, $\vec{v}$ is the coordinate-free expression for an object that, relative to a coordinate basis such as the Cartesian basis $\{\hat{\imath},\hat{\jmath},\hat{k}\}$, has the alternative expression $v^x \hat{\imath} + v^y \hat{\jmath} + v^z \hat{k}$.  
If the basis is understood, then we can get away with just referring to the coefficients  $\{v^x,v^y,v^z\}$.  
Relabeling $x,y,z\to 1,2,3$, then we can refer to either the coordinate-independent $\vec{v}$ or the coordinate coefficients $v^a$, $a=1,2,3$.
In what follows, the coordinate-free notation for four-dimensional tensorial objects will be bold, e.g. $\bm{v}$ or $\bm{F}$, and indices that range over all four space-time coordinates will be given Greek letters, e.g.\ $v^{\alpha}$ or $F_{\mu\nu}$. 
The position and ordering of indices carry meaning and are similar to the difference between row and column vectors, but indices can be raised and lowered with the space-time metric.  The metric has two subscript indices, $\bm{g}=g_{\alpha\beta}$, while its inverse has two superscript indices, $\bm{g}^{-1}=g^{\alpha\beta}$.  
The types of indices carried by an object will be given when it is introduced.
In the index-free notation I will denote index raising and lowering in two different ways.
\begin{enumerate}
\item When a single index is raised or lowered I will indicate it with the metric or its inverse, for example
\begin{equation}
  \bm{g}^{-1}\cdot\bm{F} = g^{\alpha\beta}F_{\beta\mu}, \quad \bm{F}\cdot\bm{g}^{-1} = F_{\beta\mu}g^{\mu\alpha}, \quad \bm{\mu}\cdot \bm{g} = \mu\indices{_{\mu}^{\alpha}}g_{\alpha\beta}
\end{equation}
\item When all indices are raised or lowered (e.g.\ for single-index objects) I will use the musical isomorphisms, for example
\begin{equation}
 \bm{u}^{\flat} = \bm{g}\cdot\bm{u} = g_{\alpha\beta}u^{\beta} =u_{\alpha}, \quad \bm{k}^{\sharp} = \bm{g}^{-1}\cdot\bm{k} = g^{\alpha\beta}k_{\beta} = k^{\alpha}.
\end{equation}
\end{enumerate}
Note that I have used a dot notation to indicate contraction on adjacent indices, similar to common matrix multiplication, e.g.
\begin{equation}
  \bm{F}\cdot\bm{u} = F_{\alpha\beta}u^{\beta}, \quad \bm{u}\cdot\bm{F} = u^{\alpha}F_{\alpha\beta}, \quad \bm{\mu}\cdot\bm{B} = \mu\indices{_{\mu}^{\nu}}B_{\nu}.
\end{equation}
For more complicated contractions, the index expression will be given for clarity.
Mixed index objects, such as $\bm{\mu}$ in the previous example, will be interpreted as linear endomorphisms on a vector space; e.g.\ $\bm{\mu}$ takes an element $\bm{B}$ of the cotangent space and maps it to a new element of the cotangent space $\bm{H}$.

Tensorial objects can be multiplied together in two ways such that the number of indices increases.
\begin{enumerate}
\item The tensor product $\otimes$ creates an $(m+n)$-index tensor from an $m$-index tensor and an $n$-index tensor, e.g.\
 \begin{equation}
  \bm{u}\otimes\bm{u} = u^{\alpha}u^{\beta}, \quad \bm{u}^{\flat}\otimes\bm{k}^{\sharp} = u_{\alpha}k^{\beta}.
 \end{equation}
\item The wedge product $\wedge$ is an antisymmetric version of the tensor product, with the caveat that the tensors share the same index type.  I will extend the usage of the wedge product to mixed-index objects, where the wedge shall only apply in the common exterior algebra, i.e.\ to indices of the same type, e.g.\
\begin{equation}
 \bm{u}^{\flat}\wedge\bm{E} = \bm{u}^{\flat}\otimes\bm{E} - \bm{E}\otimes\bm{u}^{\flat} = u_{\alpha}E_{\beta}-u_{\beta}E_{\alpha} = \delta^{\mu\nu}_{\alpha\beta}u_{\mu}E_{\nu}, \quad 
\end{equation}
\begin{equation}
 \bm{u}^{\flat}\wedge\bm{\delta} = \delta^{\mu\nu}_{\alpha\beta}u_{\mu}, \quad \bm{\varepsilon}\wedge\bm{u} = \delta^{\sigma\rho}_{\mu\nu}\varepsilon\indices{_{\alpha}^{\mu}}u^{\nu}.
\end{equation}
\end{enumerate}

Last, two tensors next to each other with no symbol between them will indicate contraction on all possible indices, as in
\begin{equation}
 \star\bm{F} = \star\indices{_{\alpha\beta}^{\mu\nu}}F_{\mu\nu}, \quad (\bm{\delta}\wedge\bm{u})\bm{F} = \delta_{\alpha\beta}^{\mu\nu}u^{\beta}F_{\mu\nu}
\end{equation}

\section{Tensorial electrodynamics} \label{Sec:Electrodynamics}

Returning to Maxwell's Eqs.~(\ref{Eq:Maxwell3D}), the apparent symmetry of the equations, and the fact that $\vec{E}$ and $\vec{B}$ may be obtained from the coupled set of potentials in Eqs.\ (\ref{Eq:3DPotentials}), hints at a deeper connection between the various fields.
This connection may be better illuminated from the vantage of a unified space-time and the well defined operations of differential geometry on manifolds.
The scalar and vector potentials may be combined into the four-dimensional potential 1-form $\bm{A}$, which in flat space-time is simply related to the usual scalar and vector potentials
\begin{equation}
 \bm{A}=A_{\mu} = (-\phi,\vec{A}).  
\end{equation}
It turns out that electrodynamics is particularly well suited to description by exterior calculus. Turning to flat space-time and adopting Cartesian coordinates, the exterior derivative of $\bm{A}$ can be expressed in matrix form as
\begin{equation} \label{Eq:dAMatrix}
(\ed\bm{A})_{\mu\nu}= A_{\nu,\mu}-A_{\mu,\nu} =\begin{pmatrix}
  0 & \frac{\partial \phi}{\partial x}+\frac{\partial A_x}{\partial t} & \frac{\partial \phi}{\partial y}+\frac{\partial A_y}{\partial t} & \frac{\partial \phi}{\partial z}+\frac{\partial A_z}{\partial t} \\

  -\frac{\partial \phi}{\partial x}-\frac{\partial A_x}{\partial t} & 0 & \frac{\partial A_y}{\partial x}-\frac{\partial A_x}{\partial y} & \frac{\partial A_z}{\partial x}-\frac{\partial A_x}{\partial z} \\

  -\frac{\partial \phi}{\partial y}-\frac{\partial A_y}{\partial t} & \frac{\partial A_x}{\partial y}-\frac{\partial A_y}{\partial x} & 0 & \frac{\partial A_z}{\partial y}-\frac{\partial A_y}{\partial z}  \\

  -\frac{\partial \phi}{\partial z}-\frac{\partial A_z}{\partial t} & \frac{\partial A_x}{\partial z}-\frac{\partial A_z}{\partial x} & \frac{\partial A_y}{\partial z}-\frac{\partial A_z}{\partial y} & 0 \\
 \end{pmatrix}
\end{equation}
the elements of which clearly show the relations Eqs.~(\ref{Eq:3DPotentials}) in component form.  
Making the identifications suggested by Eqs.~(\ref{Eq:3DPotentials}), we find that in the unified space-time approach the components of $\vec{E}$ and $\vec{B}$ are combined into a 2-form 
\begin{equation}
 \bm{F} = \ed \bm{A}
\end{equation}
called the \textit{field strength} tensor.
In the four-dimensional description, $\vec{E}$ and $\vec{B}$ are no longer distinct objects, but are instead merely selected components of the tensor $\bm{F}$.
The exterior derivative increases the number of tensor indices by one, so the exterior derivative of the 1-form $A_{\mu}$ is the 2-form $F_{\mu\nu}$, the coefficients of which were written as a $4\times 4$ matrix in Eq.~(\ref{Eq:dAMatrix}).
The space of 1-forms has coordinate basis $\{\ed x^{\alpha}\}$, the space of 2-forms has basis $\{\ed x^{\alpha}\wedge\ed x^{\beta}\}$, the space of 3-forms has basis $\{\ed x^{\alpha}\wedge\ed x^{\beta}\wedge \ed x^{\gamma}\}$, etc., so the matrix expression of a 2-form represents the coefficients of a linear combination of basis elements
\begin{equation} \label{Eq:FDecomposition}
 \bm{F} = \frac12 F_{\mu\nu}(\mathrm{d}x^{\mu}\wedge \mathrm{d}x^{\nu}) = E_a(\mathrm{d}x^a\wedge \mathrm{d}t) + \frac12 B_{ab}(\mathrm{d}x^a\wedge \mathrm{d}x^b).
\end{equation}
Just as the component functions used to describe a vector differ among observers according to their independent coordinate systems, so too do the component functions of $\bm{F}$ differ among observers.
Thus while one observer may only see nonzero values for $\bm{F}$ in the components labeled by $E_{a}$, another observer using different (space-time) coordinates might see nonzero components that she labels by $B_{ab}$.

The field strength tensor $\bm{F}$ encodes information about the electric field strength $E_{a}$ and the magnetic flux $B_{ab}$.
It is no coincidence that we identify the ``field'' aspect with the time-space components of $\mf$ and the ``flux'' aspect with the space-space components; the rules for integration on manifolds lead naturally to the integral forms of Faraday's law and Gauss' law for magnetism, where $\vec{B}$ has a clear interpretation as a flux through a spatial surface, as discussed in Sec.~\ref{Sec:Integration}.
In a similar manner we need to encode information about the fields $\vec{H}$ and $\vec{D}$, so it should come as no surprise that we let the \textit{excitation tensor} 
\begin{equation} \label{Eq:GDecomposition}
\bm{G} = -H_a(\mathrm{d}x^a\wedge \mathrm{d}t) + \frac12 D_{ab}(\mathrm{d}x^a\wedge \mathrm{d}x^b)
\end{equation}
encode information about the magnetic field strength $H_a$ and electric flux $D_{ab}$.

Comparing Eq.\ (\ref{Eq:GDecomposition}) with Eq.\ (\ref{Eq:FDecomposition}) it seems clear that if we want to conserve index type, then the simple vacuum relationships $\vec{D} = \vec{E}$ and $\vec{H}=\vec{B}$ need to be replaced by something else when talking about $E_a$ and $D_{ab}$ or $H_a$ and $B_{ab}$.
The 3-vector relations of Eqs.~(\ref{Eq:3VecConstitutives}) must therefore be relating the coefficients of 2-forms to those of 1-forms without the explicit use of an operation that is designed for such a task.  
In fact,  in vacuum the dual nature of these fields is related by just such an operator, called the Hodge dual,
\begin{equation}
 \bm{G} = \star\bm{F},
\end{equation}
as discussed in more detail in the next subsection.

\subsection{Differential form of {M}axwell's equations} \label{Sec:Differential}

An important property of the exterior derivative is that it is nilpotent, so $\mathrm{d}^2\bm{\omega}=0$ for any differential form $\bm{\omega}$.  
Since $\bm{F}=\mathrm{d}\bm{A}$, it immediately follows that 
\begin{equation}
 \mathrm{d}\bm{F}=0.
\end{equation}
This single equation is equivalent to the pair of homogeneous Maxwell's equations, which can seen by explicitly calculating the exterior derivative of $\ed\ed\bm{A}$ in Minkowski space-time \cite{Jackson}.

The homogeneous equation does not contain any information about the space-time, and can be thought of as a constraint equation for the fields.
Information about the space-time appears in the inhomogeneous equations, where it enters through the Hodge dual, $\star$.
The Hodge dual is defined for any pseudo-Riemannian manifold that possesses a metric and a volume form, and contains all the space-time information of the metric tensor.
In particular, the fundamental electromagnetic fields are differential forms (antisymmetric tensors).  Let $\Lambda^k T_p M$ and $\Lambda^k T^*_p M$ denote, respectively, the spaces of  alternating tensor $k$-products of tangent and cotangent spaces of manifold $M$ at point $p$.
An alternating $k$-vector field ($k$-blade) is an element of $\Lambda^k T M = \bigcup_{p\in M}\Lambda^k T_p M$, while an alternating $k$-covector field ($k$-form) is an element of  $\Lambda^k T^* M = \bigcup_{p\in M}\Lambda^k T^*_p M$.
The metric (and its inverse) provides a map between tangent and cotangent spaces by raising and lowering indices, and is thus extensible to a map $\bm{g}:\Lambda^k T M \to \Lambda^k T^* M$, while the covariant volume form (and its inverse) provides a map $\bm{\omega}:\Lambda^k T M \to \Lambda^{(m-k)} T^* M$.  The composition of these two maps defines the Hodge dual as a map $\star:\Lambda^k T M \to \Lambda^{(m-k)} T M$ or  $\star:\Lambda^k T^* M \to \Lambda^{(m-k)} T^* M$, such that 
\begin{equation}
\xymatrix{
\wedge^{(m-k)}T^*_p(M) \ar[d]_{\star} \ar[r]^{g} \ar[rd]| \omega & \ar[l] \ar[ld]| \omega \wedge^{(m-k)}T_p(M) \ar[d]^{\star} \\
\wedge^{k}T^*_p(M) \ar[u] \ar[r]_{g} \ar[ru]| \omega  & \ar[l] \ar[lu]|  \omega  \wedge^{k}T_p(M) \ar[u]}
\end{equation}
is (modulo some numerical factors) commutative.
For example, the Hodge dual of the 2-form $\bm{F}$ has the component expression
\begin{equation}
 \star\bm{F} = \frac12 (\bm{\omega}\circ \bm{g}^{-1})\bm{F} = \frac12 \sqrt{-g} \epsilon_{\alpha\beta\sigma\rho}g^{\sigma\mu}g^{\rho\nu}F_{\mu\nu}
\end{equation}
where $g = \det(\bm{g})$ and $\epsilon_{\alpha\beta\sigma\rho}$ is the completely antisymmetric Levi-Civit\`{a} symbol.
An important property of $\star$ is that 
\begin{equation} \label{Eq:HodgeSquared}
 \star\star\bm{\omega} = (-1)^{\ell} \bm{\omega}
\end{equation}
where the power depends on the dimension of the manifold and the degree of $\bm{\omega}$.  For a 2-form in a four-dimensional space-time $\ell=1$, and this is the only case we will be concerned with.

Consider now the Yang-Mills action
\begin{equation} \label{Eq:Action}
 S = \int \frac12 \bm{F}\wedge\star\bm{F} - \bm{A}\wedge\bm{J}
\end{equation}
where $\bm{J} = \star\bm{j}$ is the charge-current 3-form source.
Varying the action with respect to $\bm{A}$ provides the inhomogeneous Maxwell equation
\begin{equation}
 \ed\star\bm{F} = \bm{J}.
\end{equation}

In the presence of ponderable media, an incident field $\bm{F}$ can induce multipole moments of $\bm{J}$ that can contribute to the source term even if the total monopole (i.e. free) charge contribution to $\bm{J}$ may vanish.
In such a case the solutions to Maxwell's equations include the particular solution
\begin{equation}
 \ed\star\bm{P} = \bm{J}_{multipole}.
\end{equation} 
Then the general solution satisfies
\begin{equation}
 \ed\bm{F} = 0,
\end{equation}
\begin{equation} 
 \ed\star(\bm{F}+\bm{P}) = \ed\bm{G} = \bm{J}_{free},
\end{equation}
and we now have a generalization of Eqs.~(\ref{Eq:DEPconst}) and (\ref{Eq:BHMconst}) in the form of
\begin{equation}
 \bm{G} = \star(\bm{F} + \bm{P}).
\end{equation}
Generalizing Eq.~(\ref{Eq:PolarizationVec}), the polarization tensor $\bm{P}$ should, in principle, be expandable in terms of $\bm{F}$ and its derivatives.
Of interest here is when the polarization is linear in $\bm{F}$, whence the previous constitutive relation simplifies to
\begin{equation} \label{Eq:CovariantConstitutive}
 \bm{G} = \star\bm{\chi}\bm{F},
\end{equation}
or in indices
\begin{equation}  \label{Eq:ConstitutiveIndices}
 G\indices{_{\mu\nu}} = \star\indices{_{\mu\nu}^{\alpha\beta}}\chi\indices{_{\alpha\beta}^{\sigma\rho}}F\indices{_{\sigma\rho}}.
\end{equation}
Recall that the constitutive relation in vacuum is simply $\bm{G} = \star\bm{F}$.  The vacuum may therefore be considered a trivial linear dielectric medium for which $\bm{\chi}_{\mathrm{vac}} = id$.

\subsection{Integral form of {M}axwell's equations} \label{Sec:Integration}
We have seen the differential form version of Maxwell's equations, but in the 3-vector approach to electrodynamics one frequently wants to work with the integral form of the equations.  How do the integral equations follow from this four-dimensional approach?
Begin with the homogeneous equation $\ed\mf=0$.  Since $\ed\mf$ is a 3-form it can only be integrated over a three dimensional submanifold $\Omega\subset \mathcal{M}$
\begin{equation}
 \int_{\Omega} \ed\mf = \int_{\partial\Omega} \mf = 0,
\end{equation}
where the intermediate step follows from the generalized Stokes theorem.
The integral of the 2-form $\bm{F}$ is over $\partial\Omega$, the two-dimensional boundary of $\Omega$.  
  
\subsubsection{Faraday's law}
Consider a simple example in flat space-time and using Cartesian coordinates.  
In a curved space-time one would have to pullback the integrand to a submanifold of a chart. 
To proceed with the integration one must first choose $\Omega$, which we suppose to be the three dimensional surface consisting of the $xy$ plane extending into $t$.  
In other words, the restriction of $\ed\mf$ to $\Omega$ is $(dF)_{txy}\ed t\wedge\ed x\wedge\ed y$.  
At first glance one may suppose to write the boundary terms as $F_{tx}\ed t\wedge\ed x+F_{ty}\ed t\wedge\ed y + F_{xy}\ed x \wedge\ed y$, but special attention must be paid to the orientation of $\partial\Omega$, which must be compatible with the orientation of $\Omega$, which in turn must be compatible with the orientation of $\mathcal{M}$ (the orientation of which is given by the volume form $\star 1$).  
In particular, the prescription for integration whereby 
\begin{equation}
 \int_{\partial\Omega} F dx^1\wedge dx^2 \to \int_{\partial\Omega} F dx^1dx^2
\end{equation}
introduces a certain blindness to the orientation since
\begin{equation}
 \int_{\partial\Omega} -F dx^2\wedge dx^1 \to -\int_{\partial\Omega} F dx^1dx^2.
\end{equation}
By comparing with Eq.\ (\ref{Eq:FDecomposition}) we see that $\mf$ restricted to the boundary selects the terms 
\begin{equation}
 \left.\mf\right|_{\partial\Omega} = F_{xt}\ed x\wedge\ed t+F_{yt}\ed y\wedge\ed t + F_{xy}\ed x \wedge\ed y.
\end{equation}
The integral becomes
\begin{equation}
  \int_{\partial\Omega} \mf = \int F_{xt} dx dt + \int F_{yt} dy dt + \int F_{xy} dx dy
  \Rightarrow \int E_x dx+E_y dy = -\frac{\partial}{\partial t} \int B_z dx dy,
\end{equation}
which straightforwardly generalizes to
\begin{equation}
 \int \vec{E}\cdot d\vec{r} = -\frac{\partial}{\partial t}\int \vec{B}\cdot\hat{n}\ dA.
\end{equation}

\subsubsection{Gauss' law for magnetic fields}
This time choose a spatial volume for $\Omega$.  In this case we clearly have
\begin{equation}
 \left.\mf\right|_{\partial\Omega} = F_{xy}\ed x\wedge\ed y+F_{yz}\ed y\wedge\ed z + F_{zx}\ed z \wedge\ed x
\end{equation}
and it follows that
\begin{equation}
 \int_{\partial\Omega} \vec{B}\cdot\hat{n}\ dA = 0.
\end{equation}

A similar treatment of the inhomogeneous equation $\ed\bm{G} = \bm{J}$ provides Ampere's law and Gauss' law for electric fields.

\subsubsection{Vector potential source equations}
Integrating the homogeneous and inhomogeneous Maxwell equations over appropriate submanifolds gave us the usual integral forms of Maxwell's equations.  
But these are not the only well-known results from electrodynamics.  Where, for example, does the Biot-Savart law come from?  
The Biot-Savart law calculates $\bm{F}$ by differentiating the vector potential $\bm{A}$.  For sourceless free fields, one usually considers $\bm{A}$ as given and the calculation $\bm{F}=\mathrm{d}\bm{A}$ is straightforward.  But how does one determine $\bm{F}$ when $\bm{A}$ is dynamically sourced by some charge-current distribution $\bm{J}$?  As a starting point, one could try to determine $\bm{A}$ by integrating both sides of Maxwell's inhomogeneous equation
\begin{equation}
\int_{\Omega}\mathrm{d}\star\mathrm{d}\bm{A} = \int_{\Omega}\bm{J}.
\end{equation}
The next expected step might be
\begin{equation}
 \int_{\partial\Omega}\star\mathrm{d}\bm{A} = \int_{\Omega}\bm{J},
\end{equation}
but then we can go no further and it seems we have reached an impasse. Instead of a direct integration, $\bm{A}$ must be obtained from Maxwell's inhomogeneous equation with the assistance of Green's technique.  In the Minkowski vacuum with only free charges, this Green function approach leads to Jifemenko's equations \cite{Jackson}.

\subsection{Properties of the constitutive relation} \label{Sec:ChiProperties}
The constitutive quantity $\bm{\chi}$ may be thought of as a map from 2-forms to 2-forms, which means that it can have at most 36 independent components.
Because of this, it is possible to reformulate everything in terms of 2 six-dimensional vectors $(\vec{E},\vec{B})$ and $(\vec{D},\vec{H})$, and a $6\times 6$ constitutive matrix, and
this approach is frequently taken in the literature.
I do not follow this approach for three reasons: 1) the extraction of $\vec{E}$ and $\vec{B}$ is usually done by hand in a somewhat \textit{ad hoc} way that typically relies on a Minkowski space-time decomposition; 2) the vectors $\vec{B}$ and $\vec{D}$ obscure the 2-form nature of these fields, which becomes more apparent in the integral formulation of Maxwell's equations, where $\vec{B}$ and $\vec{D}$ appear in integrals over a 2-surface; and 3) since modern computer algebra
systems can easily handle large matrix calculations there is no computational advantage to using a six-dimensional formulation, while at the same time a preferred use of the four-dimensional differential forms formulation is consistent with
the rest of general relativity, and hence allows the possibility to incorporate the electrodynamics developed here in the standard relativistic framework.

We can demand that $\bm{\chi}$ is independently antisymmetric on its covariant and contravariant indices.  This symmetry condition reduces the number of free parameters to exactly the required 36.
There are additional symmetries with respect to which it may sometimes be useful to decompose $\mc$.  
In particular,  consider the decomposition
\begin{equation}
 \mc = \tensor[^{P}]{\mc}{} + \tensor[^{A}]{\mc}{} + \tensor[^S]{\mc}{}
\end{equation}
where
\begin{subequations}
\begin{equation}
 \tensor[^{S}]{\chi}{_{\alpha\beta\mu\nu}} = \frac12\left(\chi\indices{_{\alpha\beta\mu\nu}} - \chi\indices{_{\mu\nu\alpha\beta}}\right)
\end{equation}
\begin{equation}
  \tensor[^{A}]{\chi}{_{\alpha\beta}^{\mu\nu}} = f(\bm{x})\star\indices{_{\alpha\beta}^{\mu\nu}}.
\end{equation}
\begin{equation}
  \tensor[^{P}]{\chi}{_{\alpha\beta\mu\nu}} = \frac12\left(\chi\indices{_{\alpha\beta\mu\nu}} + \chi\indices{_{\mu\nu\alpha\beta}}\right) - \tensor[^A]{\chi}{_{\alpha\beta\mu\nu}}
\end{equation}
\end{subequations}
which ensure the symmetry properties
\begin{equation}
 \tensor[^{P}]{\chi}{_{\alpha\beta\mu\nu}}=\tensor[^{P}]{\chi}{_{\mu\nu\alpha\beta}}, \quad
 \tensor[^{A}]{\chi}{_{\alpha\beta\mu\nu}}=\tensor[^{A}]{\chi}{_{\mu\nu\alpha\beta}}, \quad 
 \tensor[^{S}]{\chi}{_{\alpha\beta\mu\nu}}=-\tensor[^{S}]{\chi}{_{\mu\nu\alpha\beta}}.
\end{equation}
Note that the symmetries here are defined with respect to the metric by lowering the second set of indices on $\bm{\chi}$, and is therefore not a fundamental decomposition of $\bm{\chi}$ on its own.
In the premetric literature these are referred to as the principal, axion, and skewon parts \cite{Hehl,Itin2008gerg,Itin2013prd}.  
Premetric electrodynamics replaces the Hodge dual appearing in the action of Eq.~(\ref{Eq:Action}) with something like $\bm{\chi}$, so one should note that their $\bm{\chi}$ has a somewhat different, although related, meaning to the $\bm{\chi}$ used here. 
In particular, one could possibly identify $(\bm{g}\otimes\bm{g})\bm{\chi}_{\mathrm{premetric}} = \sqrt{-g}\bm{\chi}$, but it should also be noted that the goals of premetric electrodynamics are somewhat different, and for electrodynamics in real media one should, strictly speaking, vary the action of Eq.~(\ref{Eq:Action}) and subsequently identify particular solutions that are linear in the homogeneous field, as done in the derivation above, rather than as an \textit{ansatz} in the action. 
The symmetric principal part $\tensor[^{P}]{\mc}{}$ has 20 components and is responsible for the usual macroscopic dielectric response parameters; 
the axion part $\tensor[^{A}]{\mc}{}$ has only one independent component; and the antisymmetric skewon part $\tensor[^{S}]{\mc}{}$ has 15 components.
Additional symmetry conditions can therefore be imposed on $\bm{\chi}$ that eliminate the skewon or axion parts,  based, for example, on thermodynamic or energy conservation arguments, or by the lack of an observed directive effect in naturally occurring stationary media \cite{Landau8,Post}.  Since here I am interested in allowing the most general medium possible, I do not pursue such a decomposition.

\section{Relation to 3-vector quantities} \label{Sec:4Dto3D}
Since most people have a more intuitive feeling for the three-dimensional vectors and medium parameters, it would be nice to explicitly show how $\bm{\chi}$ is related to the usual permeability, permittivity, and magnetoelectric couplings.
In the process, we may gain some better understanding of electrodynamics in media.
One of the remarkable things about electrodynamics is that the fields are spacelike, but since there is no global definition of time, splitting space-time into space and time depends on the choice of observers, from which it follows that the identification of electric and magnetic fields is observer dependent.
Indeed, the electric and magnetic fields are only defined relative to some observer.  When dealing with tensorial electrodynamics, the electric and magnetic field components are often extracted from $\bm{F}$ and $\bm{G}$ by hand in some (locally, at least) Minkowski space-time, but we may gain some insight by explicitly retaining the role of the observer.  
Since the covariance between observers holds in curved space-times, where they may not necessarily be related by simple Lorentz transformations, this approach retains full covariance of the theory.

\subsection{Space-time splitting}
Consider an observer moving with 4-velocity $\bm{u}$ with squared magnitude $\bm{g}(\bm{u},\bm{u})=(\bm{u}^{\flat}\cdot\bm{u})$ (or, strictly speaking, let $\bm{u}$ be a hypersurface orthogonal timelike vector field representing a family of observers). 
Relative to this observer, define the electric and magnetic 1-form fields by
\begin{equation}
 \bm{E} = (\bm{u}^{\flat}\cdot\bm{u})^{-1} \bm{u}\cdot\bm{F} \quad \text{and} \quad \bm{B} = -(\bm{u}^{\flat}\cdot\bm{u})^{-1} \bm{u}\cdot\star\bm{F}
\end{equation}
which have the index expressions
\begin{equation}
  E_{\beta} =(\bm{u}^{\flat}\cdot\bm{u})^{-1}F_{\alpha\beta}u^{\alpha} \quad \text{and} \quad B_{\beta} = -(\bm{u}^{\flat}\cdot\bm{u})^{-1}\star\indices{_{\alpha\beta}^{\mu\nu}}F_{\mu\nu}u^{\alpha}.
\end{equation}
Notice that by the antisymmetry of $\bm{F}$, we can express $\bm{E}$ and $\bm{B}$ with the somewhat more symmetric expressions
\begin{equation} \label{Eq:SymEB}
 \bm{E} = -\frac12 (\bm{u}^{\flat}\cdot\bm{u})^{-1} (\bm{\delta}\wedge\bm{u})\bm{F} \quad \text{and} \quad \bm{B} = \frac12 (\bm{u}^{\flat}\cdot\bm{u})^{-1} (\bm{\delta}\wedge\bm{u})\star\bm{F}.
\end{equation}
Although these last expressions are somewhat tautological, they will be useful shortly.
It is immediately clear by the antisymmetry properties of $\bm{F}$ and $\star$ that $\bm{u}\cdot\bm{E} = \bm{u}\cdot\bm{B} = 0$.  
Thus the 1-form fields $\bm{E}$ and $\bm{B}$ are orthogonal to $\bm{u}$.
We can reconstruct $\bm{F}$ by
\begin{equation}
 \bm{F} = \bm{u}^{\flat}\wedge\bm{E} + \star(\bm{u}^{\flat}\wedge\bm{B}),
\end{equation}
which has the index expression
\begin{equation}
 F_{\mu\nu} = 2\left( u_{[\mu}E_{\nu]} + \star\indices{_{\mu\nu}^{\alpha\beta}}u_{[\alpha}B_{\beta]} \right).
\end{equation}
The dual fields have a similar structure.  According to observer $\bm{u}$, they are
\begin{equation} \label{Eq:DH}
 \bm{D} = -(\bm{u}^{\flat}\cdot\bm{u})^{-1} \bm{u}\cdot\star\bm{G} \quad \text{and} \quad \bm{H} = -(\bm{u}^{\flat}\cdot\bm{u})^{-1}\bm{u}\cdot\bm{G},
\end{equation}
while
\begin{equation}
 \bm{G} = -\bm{u}^{\flat}\wedge\bm{H} + \star(\bm{u}^{\flat}\wedge\bm{D}).
\end{equation}
Clearly, $\bm{D}$ and $\bm{H}$ are also orthogonal to $\bm{u}$, $\bm{u}\cdot\bm{D} = \bm{u}\cdot\bm{H} =0$.

\subsection{Obtaining usual parameters from $\bm{\chi}$}
We have seen that within linear dielectric media, the constitutive relation is $\bm{G}=\star\bm{\chi}\bm{F}$.  We can use this to connect to the usual concepts of permittivity, permeability, and magnetoelectric couplings in the following way.
From Eq.~(\ref{Eq:DH}) one finds
\begin{equation}
 \begin{aligned} 
  \bm{D} & = -(\bm{u}^{\flat}\cdot\bm{u})^{-1}\bm{u}\cdot \star\bm{G} = -(\bm{u}^{\flat}\cdot\bm{u})^{-1}\bm{u}\cdot(\star\star\bm{\chi}\bm{F}) = (\bm{u}^{\flat}\cdot\bm{u})^{-1}\bm{u}\cdot\bm{\chi}\bm{F} \\
         & = (\bm{u}^{\flat}\cdot\bm{u})^{-1} \bm{u}\cdot \bm{\chi}\left(\bm{u}^{\flat}\wedge\bm{E}+\star(\bm{u}^{\flat}\wedge\bm{B})\right) \\
         & = -(\bm{u}^{\flat}\cdot\bm{u})^{-1} \left[ 2(\bm{u}\cdot\bm{\chi}\cdot\bm{u}^{\flat})\cdot\bm{E} + 2(\bm{u}\cdot\bm{\chi}\star\cdot\bm{u}^{\flat})\cdot\bm{B} \right]\\
         & = \bm{\varepsilon}^{c}\cdot \bm{E} +  \tensor[^b]{\bm{\gamma}}{}^c \cdot \bm{B}
 \end{aligned} 
\end{equation}
where the third line follows from the antisymmetry of $\bm{\chi}$ in its second set of indices.
Similarly, for $\bm{H}$ one finds
\begin{equation}
 \begin{aligned}
  \bm{H} & = -(\bm{u}^{\flat}\cdot\bm{u})^{-1} \bm{u}\cdot\bm{G} = -(\bm{u}^{\flat}\cdot\bm{u})^{-1} \bm{u}\cdot\star\bm{\chi}\bm{F} \\
         & =  -(\bm{u}^{\flat}\cdot\bm{u})^{-1} \bm{u}\cdot\star\bm{\chi}\left(\bm{u}^{\flat}\wedge\bm{E} + \star(\bm{u}^{\flat}\wedge\bm{B}) \right) \\
         & = (\bm{u}^{\flat}\cdot\bm{u})^{-1}\left[ 2(\bm{u}\cdot\star\chi\cdot\bm{u}^{\flat})\cdot \bm{E} + 2(\bm{u}\cdot\star\chi\star\cdot\bm{u}^{\flat})\cdot\bm{B}\right] \\
         & = \tensor[^e]{\bm{\gamma}}{}^c \cdot \bm{E} + \bm{\xi}\cdot \bm{B}.
 \end{aligned}
\end{equation}
To summarize, we now have the parameters
\begin{subequations} \label{Eq:4Dparameters}
 \begin{equation}
  \begin{aligned}
   \bm{\varepsilon}^{c} & = -2(\bm{u}^{\flat}\cdot\bm{u})^{-1} (\bm{u}\cdot\bm{\chi}\cdot\bm{u}^{\flat}) \\
   (\varepsilon^c)\indices{_{\beta}^{\mu}} & = -2(u^{\lambda}u_{\lambda})^{-1} u^{\alpha}\chi\indices{_{\alpha\beta}^{\mu\nu}}u_{\nu} 
  \end{aligned}
 \end{equation}
 \begin{equation}
  \begin{aligned} 
   \bm{\xi} & = 2(\bm{u}^{\flat}\cdot\bm{u})^{-1}(\bm{u}\cdot\star\chi\star\cdot\bm{u}^{\flat}) \\ 
   \xi\indices{_{\beta}^{\mu}} & =  2 (u^{\lambda}u_{\lambda})^{-1} u^{\alpha} \star\indices{_{\alpha\beta}^{\lambda\kappa}} \chi\indices{_{\lambda\kappa}^{\sigma\rho}} \star\indices{_{\sigma\rho}^{\mu\nu}}u_{\nu}
  \end{aligned} 
 \end{equation}
 \begin{equation}
  \begin{aligned}
    \tensor[^b]{\bm{\gamma}}{}^c & = -2(\bm{u}^{\flat}\cdot\bm{u})^{-1} (\bm{u}\cdot\bm{\chi}\star\cdot\bm{u}^{\flat}) \\
    (\tensor[^b]{\gamma}{}^c)\indices{_{\beta}^{\mu}} &  = - 2(u^{\lambda}u_{\lambda})^{-1} u^{\alpha}\chi\indices{_{\alpha\beta}^{\lambda\kappa}}\star\indices{_{\lambda\kappa}^{\mu\nu}}u_{\nu}
  \end{aligned}
 \end{equation}
 \begin{equation}
  \begin{aligned}
    \tensor[^e]{\bm{\gamma}}{}^c & = 2(\bm{u}^{\flat}\cdot\bm{u})^{-1} (\bm{u}\cdot\star\chi\cdot\bm{u}^{\flat}) \\
     (\tensor[^e]{\gamma}{}^c)\indices{_{\beta}^{\mu}} & =  2(u^{\lambda}u_{\lambda})^{-1} u^{\alpha} \star\indices{_{\alpha\beta}^{\lambda\kappa}} \chi\indices{_{\lambda\kappa}^{\mu\nu}}u_{\nu}.
  \end{aligned}
 \end{equation}
\end{subequations}
Note that Eqs.~(\ref{Eq:4Dparameters}) are space-time expressions, so each of these is a $4\times 4$ tensor rather than a $3\times 3$ matrix.
However, since $\bm{u}\cdot\bm{D} = 0$ one can readily see that we  must have 
\begin{equation} \label{Eq:LeftOrthogonal}
\bm{u}\cdot\bm{\varepsilon}^c=0, \quad u^{\alpha}(\varepsilon^c)\indices{_{\alpha}^{\beta}}=0
\end{equation}
 and similar for the other material tensors.
Since $\bm{u}$ is orthogonal to $\bm{E}$ and $\bm{B}$ in the domain of each material tensor, it follows that each of $\bm{\varepsilon}^c$, $\bm{\xi}$, $\tensor[^b]{\bm{\gamma}}{}^c$, and $\tensor[^e]{\bm{\gamma}}{}^c$ are linear automorphisms of the three-dimensional subspace of the cotangent bundle $T^*(M)$ orthogonal to $\bm{u}$; thus it is also true that
\begin{equation} \label{Eq:RightOrthogonal}
 \bm{\varepsilon}^c\cdot\bm{u}^{\flat} = 0, \quad (\varepsilon^c)\indices{_{\alpha}^{\beta}}u_{\beta} =0
\end{equation}
and similar for $\bm{\xi}$, $\tensor[^b]{\bm{\gamma}}{}^c$, and $\tensor[^e]{\bm{\gamma}}{}^c$. 
For example, in a local Cartesian frame comoving with the observer, such that $u^{\alpha}=(1,0,0,0)$, one would find
\begin{equation} \label{Eq:CartesianEpsilon}
 (\varepsilon^c)\indices{_{\alpha}^{\mu}} =
 \begin{pmatrix}
  0 & 0 & 0 & 0 \\ 
  0 & \varepsilon^{cx}_{x} & \varepsilon^{cy}_{x} & \varepsilon^{cz}_{x} \\
  0 & \varepsilon^{cx}_{y} & \varepsilon^{cy}_{y} & \varepsilon^{cz}_{y} \\
  0 & \varepsilon^{cx}_{z} & \varepsilon^{cy}_{z} & \varepsilon^{cz}_{z} \\
 \end{pmatrix}. 
\end{equation}
In this comoving picture (equivalently the observer is ``at rest''), the 3-vector \textit{matrix} relations are recovered simply by taking the purely spatial part of the \textit{tensor} relations in a local Cartesian frame.

Since the timelike vector field $\bm{u}$ defines a spacelike foliation of the manifold, and electrodynamics takes place in the purely spatial three-dimensional subspace orthogonal to $\bm{u}$, then it is useful to define an operator
\begin{equation}
 \bm{h} = \bm{\delta} - \frac{\bm{u}^{\flat}\otimes\bm{u}}{(\bm{u}^{\flat}\cdot\bm{u})}, \quad h_{\alpha}^{\beta} = \delta_{\alpha}^{\beta} - \frac{u_{\alpha}u^{\beta}}{u_{\mu}u^{\mu}}
\end{equation} 
that projects out the purely spatial part of a vector or 1-form, relative to $\bm{u}$. 
In other words, $\bm{h}$ annihilates any component proportional to $\bm{u}$.
Note that $\bm{h}\cdot\bm{h} = \bm{h}$ and that the Kronecker delta projects to $\bm{h}\cdot\bm{\delta}\cdot\bm{h} = \bm{h}$, so $\bm{h}$ serves as the Kronecker delta on the subspace orthogonal to $\bm{u}$.
Consequently, in this four-dimensional notation, the vacuum permeability and permittivity are $\bm{\mu}= \bm{\varepsilon} = \bm{h}$.
By Eqs.~(\ref{Eq:LeftOrthogonal}) and (\ref{Eq:RightOrthogonal}) we can see that since $\bm{\varepsilon}^c$, $\bm{\xi}$, $\tensor[^b]{\bm{\gamma}}{}^c$, and $\tensor[^e]{\bm{\gamma}}{}^c$ are already orthogonal to $\bm{u}$, then $\bm{h}\cdot\bm{\varepsilon}^c\cdot\bm{h} =  \bm{\varepsilon}^c$, etc.

\subsection{Reconstruction of $\bm{\chi}$ from transverse parameters}

Now it is possible to reconstruct $\bm{\chi}$ out of the transverse parameters.
Acting on both sides of the constitutive relation $\bm{G} = \star\bm{\chi}\bm{F}$ with $\star$, one finds 

\begin{equation} 
 \begin{aligned}
  \bm{\chi}\bm{F} = -\star\bm{G}   
   & = \star(\bm{u}^{\flat}\wedge\bm{H}) + \bm{u}^{\flat}\wedge\bm{D} \\
   & = \star\left( \bm{u}^{\flat} \wedge (\tensor[^e]{\bm{\gamma}}{}^c \cdot \bm{E} + \bm{\xi}\cdot \bm{B})\right) + \bm{u}^{\flat} \wedge (\bm{\varepsilon}^{c}\cdot \bm{E} +  \tensor[^b]{\bm{\gamma}}{}^c \cdot \bm{B})  \\
   & = \star( \bm{u}^{\flat} \wedge \tensor[^e]{\bm{\gamma}}{}^c )\cdot\bm{E} + \star( \bm{u}^{\flat} \wedge \bm{\xi})\cdot \bm{B} +  (\bm{u}^{\flat} \wedge \bm{\varepsilon}^{c}) \cdot \bm{E} + (\bm{u}^{\flat} \wedge  \tensor[^b]{\bm{\gamma}}{}^c)\cdot \bm{B}.
 \end{aligned}
\end{equation}
Using Eqs.~(\ref{Eq:SymEB}) for $\bm{E}$ and $\bm{B}$, this becomes
\begin{equation} 
   \bm{\chi}\bm{F} 
  = \frac12 (\bm{u}^{\flat}\cdot\bm{u})^{-1} \left[ -\star( \bm{u}^{\flat} \wedge \tensor[^e]{\bm{\gamma}}{}^c \wedge\bm{u}) + \star( \bm{u}^{\flat} \wedge \bm{\xi}\wedge\bm{u})\star -  (\bm{u}^{\flat} \wedge  \bm{\varepsilon}^{c} \wedge \bm{u}) + (\bm{u}^{\flat} \wedge  \tensor[^b]{\bm{\gamma}}{}^c\wedge\bm{u})\star\right]\bm{F}.\\
\end{equation}
Rearranging terms a little, the sought after identity is clearly
\begin{equation} \label{Eq:ChiWedges}
 \bm{\chi} = \frac12 (\bm{u}^{\flat}\cdot\bm{u})^{-1} \left[-  (\bm{u}^{\flat} \wedge  \bm{\varepsilon}^{c} \wedge \bm{u}) + \star( \bm{u}^{\flat} \wedge \bm{\xi}\wedge\bm{u})\star   - \star( \bm{u}^{\flat} \wedge \tensor[^e]{\bm{\gamma}}{}^c \wedge\bm{u})  + (\bm{u}^{\flat} \wedge  \tensor[^b]{\bm{\gamma}}{}^c\wedge\bm{u})\star\right].
\end{equation}
The corresponding index expression is
\begin{equation} \label{Eq:ChiIndices}
 \chi\indices{_{\gamma\delta}^{\sigma\rho}} = 2(u^{\lambda}u_{\lambda})^{-1} \left[  - u_{[\gamma}(\varepsilon^c)\indices{_{\delta]}^{[\sigma}}u^{\rho]}
 + \star\indices{_{\gamma\delta}^{\alpha\beta}}u\indices{_{[\alpha}}(\mu^{-1})\indices{_{\beta]}^{[\mu}}u^{\nu]}\star\indices{_{\mu\nu}^{\sigma\rho}}
 - \star\indices{_{\gamma\delta}^{\alpha\beta}}u_{[\alpha}(\tensor[^e]{\gamma}{}^c)\indices{_{\beta]}^{[\sigma}}u^{\rho]}
 + u_{[\gamma}(\tensor[^b]{\gamma}{}^c)\indices{_{\delta]}^{[\mu}}u^{\nu]}\star\indices{_{\mu\nu}^{\sigma\rho}}
 \right].
\end{equation}

So far, I have related $\mc$ to the set of transverse components $\{\bm{\varepsilon}^c,\bm{\xi}, \tensor[^e]{\bm{\gamma}}{}^c,\tensor[^b]{\bm{\gamma}}{}^c\}$, corresponding to the pair of constitutive relations
\begin{equation}
 \begin{aligned}
   \bm{D} & = \bm{\varepsilon}^c \cdot \bm{E} + \tensor[^b]{\bm{\gamma}}{}^c \cdot \bm{B}, \\
   \bm{H} & = \bm{\xi} \cdot \bm{B} + \tensor[^e]{\bm{\gamma}}{}^c \cdot \bm{E}
 \end{aligned}
\end{equation}
that relate the set of fields $\{\bm{D},\bm{H}\}$ to the set $\{\bm{E},\bm{B}\}$.
Although these relations follow most naturally from the four-dimensional constitutive relation $\bm{G} = \star\mc\bm{F}$, they are not the only way of expressing the transverse constitutive relations.
In traditional electrodynamics, the constitutive relations usually relate the set of fields $\{\bm{D},\bm{B}\}$ to the set $\{\bm{E},\bm{H}\}$ via
\begin{equation} \label{Eq:TradConstitutives}
 \begin{aligned}
   \bm{D} & = \bm{\varepsilon} \cdot \bm{E} + \tensor[^h]{\bm{\gamma}}{} \cdot \bm{H}, \\
   \bm{B} & = \bm{\mu} \cdot \bm{H} + \tensor[^e]{\bm{\gamma}}{} \cdot \bm{E}.
 \end{aligned}
\end{equation}
It is easy to see that these two representations are related by
\begin{equation} \label{Eq:ConstitutiveRelations}
  \bm{\xi} = \bar{\bm{\mu}}, \quad \bm{\varepsilon}^c = \bm{\varepsilon} - \tensor[^h]{\bm{\gamma}}{}\cdot \bm{\xi} \cdot \tensor[^e]{\bm{\gamma}}{}   , \quad \tensor[^e]{\bm{\gamma}}{}^c = -\bm{\xi} \cdot  \tensor[^e]{\bm{\gamma}}{}, \quad  \tensor[^b]{\bm{\gamma}}{}^c = \tensor[^h]{\bm{\gamma}}{}\cdot \bm{\xi}.
\end{equation}
Of course, the matrix displayed in Eq.~(\ref{Eq:CartesianEpsilon}) clearly does not have an inverse, and neither does $\bm{\xi}$.
However, the bar notation on $\bar{\bm{\mu}}$ denotes an inverse restricted to the three-dimensional transverse subspace.
Since any of the transverse medium parameters, such as $\bm{\mu}$, are automorphisms of the transverse subspace, then an inverse is well defined by the requirement that
\begin{equation}
 \bm{b}\cdot\bar{\bm{b}} = \bar{\bm{b}}\cdot\bm{b} = \bm{h}
\end{equation}
for any transverse map $\bm{b}$.
The restricted inverse is found from
\begin{equation} \label{Eq:3DInverse}
 \bar{b}\indices{_{\alpha}^{\beta}} =  \frac{3h_{\alpha \alpha_1 \alpha_2}^{\beta \beta_1 \beta_2} (b\indices{_{\beta_1}^{\alpha_1}})(b\indices{_{\beta_2}^{\alpha_2}})}{h_{\alpha_1 \alpha_2 \alpha_3}^{\beta_1 \beta_2 \beta_3} (b\indices{_{\beta_1}^{\alpha_1}})(b\indices{_{\beta_2}^{\alpha_2}})(b\indices{_{\beta_3}^{\alpha_3}})}
\end{equation}
where 
\begin{equation} \label{Eq:hGeneralized}
  h_{\alpha \alpha_1 \alpha_2}^{\beta \beta_1 \beta_2} = (\bm{u}^{\flat}\cdot\bm{u})^{-1} u^{\mu}u_{\nu} \delta_{\mu\alpha \alpha_1 \alpha_2}^{\nu\beta \beta_1 \beta_2}
\end{equation}
is the generalized Kronecker symbol on the transverse subspace.

The traditional representation of Eqs.~(\ref{Eq:TradConstitutives}) allows Eq.~(\ref{Eq:ChiWedges}) for $\bm{\chi}$ to be recast as
\begin{equation}
 \mc = \frac12 (\bm{u}^{\flat}\cdot\bm{u})^{-1} \left[-  (\bm{u}^{\flat} \wedge  \bm{\varepsilon} \wedge \bm{u}) + \star( \bm{u}^{\flat} \wedge \bar{\bm{\mu}}\wedge\bm{u})\star   + \star( \bm{u}^{\flat} \wedge \bar{\bm{\mu}}\cdot \tensor[^e]{\bm{\gamma}}{} \wedge\bm{u})  + (\bm{u}^{\flat} \wedge  \tensor[^h]{\bm{\gamma}}{}\cdot \bar{\bm{\mu}}\wedge\bm{u})\star +  (\bm{u}^{\flat}\wedge \tensor[^h]{\bm{\gamma}}{}\cdot \bar{\bm{\mu}} \cdot \tensor[^e]{\bm{\gamma}}{} \wedge \bm{u})  \right].
\end{equation}
This expression can actually be factored to
\begin{equation} \label{Eq:ChiTradParameters}
 \mc = \frac12 (\bm{u}^{\flat}\cdot\bm{u})^{-1} \left[-  (\bm{u}^{\flat} \wedge  \bm{\varepsilon} \wedge \bm{u}) + [ \star( \bm{u}^{\flat} \wedge \bm{h}) + \bm{u}^{\flat} \wedge  \tensor[^h]{\bm{\gamma}}{} ]\cdot  \bar{\bm{\mu}} \cdot [ (\bm{h}\wedge\bm{u})\star   +  \tensor[^e]{\bm{\gamma}}{} \wedge\bm{u}]  \right],
\end{equation}
which will be quite useful for obtaining the optical metric in the next sections.

\section{Geometric optics limit} \label{Sec:GeometricOptics}
The geometric optics limit of the wave equation for light is standard lore \cite{BornWolf}, and the presentation here essentially follows the standard method, albeit in an index-free tensorial notation.
The wave propagation of light is described by a second order equation, but Maxwell's equations provide two first order equations.
The simplest derivation of geometric optics is to first use $\bm{F}=\ed\bm{A}$ to write the inhomogeneous Maxwell equation as
\begin{equation}
 \ed\star \ed \bm{A} = \mJ
\end{equation}
and show that this is a wave equation for the 1-form potential $\bm{A}$.
Operating on both sides of this equation with the Hodge dual, it may be rewritten in terms of the codifferential of a $k$-form on an $m$-dimensional space-time manifold
\begin{equation}
 \delta = (-1)^{k(m+1)-1}\star\ed\star
\end{equation}
as
\begin{equation} \label{Eq:InhomogA}
 \delta\ed\bm{A} = -\star\bm{J}.
\end{equation}
While the exterior derivative increases the degree of a differential form by one, e.g.\ $\ed$ of a 1-form results in a 2-form, the codifferential decreases the degree by one;
so while $\ed\bm{A}$ is a 2-form, $\delta \ed\bm{A}$ is a 1-form.

Propagating fields must satisfy a hyperbolic partial differential equation, e.g.\ a wave equation.  On curved manifolds, the hyperbolic operator on differential forms is the Laplace-de Rham operator \cite{Frankel}
\begin{equation}
 \Delta = \delta \ed + \ed \delta.
\end{equation}
Adding $\ed\delta\bm{A}$ to both sides of Eq.~(\ref{Eq:InhomogA}) we have the gauge-independent wave equation
\begin{equation}
 \Delta\bm{A} = (\delta\ed + \ed\delta)\bm{A} = -\star\bm{J} + \ed\delta\bm{A}.
\end{equation}
On the other hand, fixing the gauge to the Lorenz gauge
\begin{equation}
 \delta\bm{A} = 0
\end{equation}
shows that Maxwell's inhomogeneous equation serves as a wave equation for $\bm{A}$ in the Lorenz gauge, even in media with bound source contributions.
Separating out the dipole contribution of $\bm{J}$, the form of Maxwell's equation
\begin{equation}
 \star \ed \star \mc \ed \bm{A} = \delta\chi\ed \bm{A} = 0
\end{equation}
does indeed serve as the wave equation for $\bm{A}$ in neutral, linear macroscopic media.

Since we are now dealing with a wave equation for $\bm{A}$, we can use a Jeffreys-Wentzel-Kramers-Brillouin (JWKB) type of approximation by assuming a plane wave solution of the form
\begin{equation} \label{Eq:PlaneWave}
 \bm{A} = \hat{\bm{A}}(x^{\mu})e^{-(i\lambda)^{-1} S(x^{\mu})}
\end{equation}
and retaining the leading order terms in the limit $\lambda\to 0$ \cite{BornWolf}.
In this scheme the amplitude $\hat{\bm{A}}$ is slowly varying compared to the phase function $S$ \cite{Post}.
Plugging the JWKB solution into the Lorenz gauge condition, one finds
\begin{equation}
  \delta\bm{A} = \delta\left(  \hat{\bm{A}}e^{-(i\lambda)^{-1} S} \right) = (i\lambda)^{-1} e^{-(i\lambda)^{-1} S} \left((i\lambda)\delta\hat{\bm{A}} - \bm{g}(\ed S,\hat{\bm{A}})\right) = 0. 
\end{equation}
Keeping the leading order term in the limit $\lambda\to 0$ leaves
\begin{equation} \label{Eq:kAorthogonality}
 \bm{g}(\hat{\bm{A}}, \ed S) = 0.
\end{equation}
The $1$-form $\bm{k} = \ed S$, or $k_{\nu} = \partial_{\nu}S$, is the wave covector, and Eq.~(\ref{Eq:kAorthogonality}) is the usual result that the Lorenz condition requires $\hat{\bm{A}}$ to be orthogonal to the wave vector.
Next, calculating
\begin{equation}
 \begin{aligned} 
  \delta\bm{\chi}\ed \bm{A} 
  & = (i\lambda)^{-1} \delta\left[ e^{-(i\lambda)^{-1}S}\bm{\chi}[(i\lambda)\ed\hat{\bm{A}} + \hat{\bm{A}}\wedge \ed S]\right] \\
  & =  (i\lambda)^{-2}  e^{-(i\lambda)^{-1}S} \left[ -\bm{g}\left(\ed S, \bm{\chi}[(i\lambda)\ed\hat{\bm{A}}  + \hat{\bm{A}}\wedge \ed S] \right) + (i\lambda) \delta\bm{\chi}[(i\lambda)\ed\hat{\bm{A}}  + \hat{\bm{A}}\wedge \ed S]  \right] = 0
 \end{aligned}
\end{equation}
and taking the limit $\lambda\to 0$, one finds
\begin{equation}
 -\bm{g}\left(\ed S, \bm{\chi}(\hat{\bm{A}}\wedge \ed S)\right) = 0.
\end{equation} 
In terms of $\bm{k}$, and exploiting the antisymmetry properties of $\bm{\chi}$, this may be written as
\begin{equation} \label{Eq:Eikonal}
 -(\bm{k}^{\sharp}\cdot\bm{\chi}\cdot\bm{k})\cdot \hat{\bm{A}} =  0,
\end{equation}
or with index notation as
\begin{equation}
 -g^{\alpha\sigma}\chi\indices{_{\alpha\beta}^{\mu\nu}}k_{\sigma}k_{\nu} \hat{A}_{\mu} = 0.
\end{equation}

Thinking of 
\begin{equation}
 \bm{X} = -\bm{k}^{\sharp}\cdot\bm{\chi}\cdot\bm{k}
\end{equation}
as a $4\times 4$ matrix,  it may be observed that the existence of a nontrivial solution to Eq.~(\ref{Eq:Eikonal}) requires
\begin{equation}
\det(\bm{X}) = 0.
\end{equation}
In fact, this condition is satisfied identically.
By the antisymmetry of the second set of indices on $\mc$, $\hat{\bm{A}}\propto \bm{k}$ is already a trivial solution, so any nontrivial solution resides in the three-dimensional subspace orthogonal to $\bm{k}$, meaning the matrix is effectively only three dimensional.
There are some different methods for dealing with this (see, for example, Ref.\ \cite{Hehl2002ijmpa}); I follow a purely algebraic argument based on the classical adjugate matrix $\mathrm{adj}(\bm{X})$ (used similarly by Refs.~\cite{Itin2009jpa,Thompson2017prd,Schuster2017prd}).
The adjugate is defined such that
\begin{equation}
 \bm{X} \mathrm{adj}(\bm{X}) = \det(\bm{X})\bm{I}
\end{equation}
and is closely related to the inverse; for if $\bm{X}$ is invertible, then $\mathrm{adj}(\bm{X}) \propto \bm{X}^{-1}$, but $\mathrm{adj}(\bm{X})$ is defined even if $\bm{X}^{-1}$ does not exist.
Since $\det(\bm{X}) = 0$ identically, then it must be true that $\bm{X}\mathrm{adj}(\bm{X}) = 0$.  Since $\bm{X}$ is nonzero and arbitrary, the subsidiary condition
\begin{equation}
 \mathrm{adj}(\bm{X}) = 0
\end{equation}
must be satisfied.
Although this is a matrix condition, I show below that
\begin{equation} \label{Eq:P_as_adjX}
 \mathrm{adj}(\bm{X}) = P (\bm{k}\otimes\bm{k}^{\sharp})
\end{equation}
where $P$ is a scalar polynomial of fourth order in $\bm{k}$.  Therefore, the condition for nontrivial solutions to Eq.\ (\ref{Eq:Eikonal}) reduces to the scalar condition
\begin{equation}
P = 0.
\end{equation}

To see this, begin by inserting some extra Kronecker $\bm{h}$ into the expression for $\mc$ given in Eq.~(\ref{Eq:ChiTradParameters}) to write 
\begin{equation} 
 \mc = \frac12 (\bm{u}^{\flat}\cdot\bm{u})^{-1} \left[-  (\bm{u}^{\flat}\wedge\bm{h})\cdot \bm{\varepsilon} \cdot (\bm{h}\wedge \bm{u}) + [ \star( \bm{u}^{\flat} \wedge \bm{h}) + (\bm{u}^{\flat} \wedge \bm{h})    \cdot \tensor[^h]{\bm{\gamma}}{} ]\cdot  \bar{\bm{\mu}} \cdot [ (\bm{h}\wedge\bm{u})\star   +  \tensor[^e]{\bm{\gamma}}{}\cdot(\bm{h} \wedge\bm{u})]  \right].
\end{equation}
Next, with this expression for $\mc$, compute
\begin{equation} \label{Eq:Intermediate}
 \bm{X} = -\bm{k}^{\sharp}\cdot\mc\cdot\bm{k} = -\frac12(\bm{u}^{\flat}\cdot\bm{u})^{-1}\left[ -\bm{K}\cdot\bm{\varepsilon}\cdot\bm{L} + [ \bm{M} + \bm{K} \cdot \tensor[^h]{\bm{\gamma}}{} ]\cdot  \bar{\bm{\mu}} \cdot [\bm{N}  +  \tensor[^e]{\bm{\gamma}}{}\cdot \bm{L}]  \right]
\end{equation}
where
\begin{equation}
 \bm{K} = \bm{k}^{\sharp}\cdot(\bm{u}^{\flat} \wedge \bm{h}), \quad K\indices{_{\delta}^{\beta}} = k^{\gamma}u_{\alpha}h_{\kappa}^{\beta}\delta^{\alpha\kappa}_{\gamma\delta},
\end{equation}
\begin{equation}
\bm{L} = (\bm{h}\wedge \bm{u})\cdot \bm{k}, \quad L\indices{_{\mu}^{\sigma}} =  \delta^{\sigma\rho}_{\kappa\nu}h_{\mu}^{\kappa}u^{\nu}k_{\rho},
\end{equation}
\begin{equation} \label{Eq:M}
 \bm{M} = \bm{k}^{\sharp}\cdot\star( \bm{u}^{\flat} \wedge \bm{h}), \quad  M\indices{_{\delta}^{\beta}} = -2 k^{\gamma}\star\indices{_{\gamma\delta}^{\beta\alpha}}u\indices{_{\alpha}},
\end{equation}
\begin{equation} \label{Eq:N}
 \bm{N} = (\bm{h}\wedge\bm{u})\star \cdot\ \bm{k}, \quad N\indices{_{\mu}^{\sigma}} = -2 u^{\nu}\star\indices{_{\nu\mu}^{\sigma\rho}}k_{\rho}.
\end{equation}
Note that these tensors only contain information about the observer $\bm{u}$, the wave covector $\bm{k}$, and the space-time metric (both in the $\star$ and in the musical isomorphisms that raise and lower indices on $\bm{u}$ and $\bm{k}$).  Also note that the tensors $\bm{M}$ and $\bm{N}$ are eigentensors of $\bm{K}$ and $\bm{L}$.  In particular, one may verify that
\begin{equation}
 \bm{K}\cdot\bm{M} = (\bm{k}\cdot\bm{u}) \bm{M}, \quad \mbox{and} \quad \bm{N}\cdot\bm{L} = (\bm{k}\cdot\bm{u}) \bm{N},
\end{equation}
which allows the revision of Eq.~(\ref{Eq:Intermediate}) to
\begin{equation}
 \bm{X} = -\frac12(\bm{u}^{\flat}\cdot\bm{u})^{-1} \bm{K}\cdot\left[ -\bm{\varepsilon} + [ (\bm{k}\cdot\bm{u})^{-1}\bm{M} + \tensor[^h]{\bm{\gamma}}{} ]\cdot  \bar{\bm{\mu}} \cdot [(\bm{k}\cdot\bm{u})^{-1}\bm{N}  +  \tensor[^e]{\bm{\gamma}}{}] \right]\cdot \bm{L}.
\end{equation}
By the properties of the adjugate, we have
\begin{equation}
 \mathrm{adj}(\bm{X}) = -\frac18 (\bm{u}^{\flat}\cdot\bm{u})^{-3}\mathrm{adj}(\bm{L}).\mathrm{adj}\left( -\bm{\varepsilon} + [ (\bm{k}\cdot\bm{u})^{-1}\bm{M} + \tensor[^h]{\bm{\gamma}}{} ]\cdot  \bar{\bm{\mu}} \cdot [(\bm{k}\cdot\bm{u})^{-1}\bm{N}  +  \tensor[^e]{\bm{\gamma}}{}] \right). \mathrm{adj}(\bm{K}).
\end{equation}
One may readily verify that
\begin{equation}
\mathrm{adj}(\bm{L}) =   (\bm{k}\cdot\bm{u})^2 \bm{k}\otimes \bm{u}, \quad \mbox{and} \quad \mathrm{adj}(\bm{K}) =  (\bm{k}\cdot\bm{u})^2 \bm{u}^{\flat} \otimes \bm{k}^{\sharp},
\end{equation}
so that
\begin{equation}
 \mathrm{adj}(\bm{X}) = -\frac18 (\bm{u}^{\flat}\cdot\bm{u})^{-3}(\bm{k}\cdot\bm{u})^4 \left(\bm{u}.\mathrm{adj}\left( -\bm{\varepsilon} + [ (\bm{k}\cdot\bm{u})^{-1}\bm{M} + \tensor[^h]{\bm{\gamma}}{} ]\cdot  \bar{\bm{\mu}} \cdot [(\bm{k}\cdot\bm{u})^{-1}\bm{N}  +  \tensor[^e]{\bm{\gamma}}{}] \right). \bm{u}^{\flat}\right) (\bm{k}\otimes\bm{k}^{\sharp}).
\end{equation}
Since each tensor in $\mathrm{adj}(\cdots)$ above is transverse, then the quantity $\bm{u}\cdot\mathrm{adj}(\cdots)\cdot\bm{u}^{\flat}$ is equivalent to the determinant restricted to the subspace orthogonal to $\bm{u}$.  This enables the application of Sylvester's determinant theorem to bring out $\bm{\varepsilon}$,
 \begin{equation}
  \mathrm{adj}(\bm{X}) = -\frac18 (\bm{u}^{\flat}\cdot\bm{u})^{-3}(\bm{k}\cdot\bm{u})^4 \left(\bm{u}\cdot\mathrm{adj}(\bm{\varepsilon})\cdot\mathrm{adj}\left(\bm{Q} \right)\cdot \bm{u}^{\flat}\right) (\bm{k}\otimes\bm{k}^{\sharp}),
 \end{equation}
with
\begin{equation} \label{Eq:Q}
 \bm{Q} = -\bm{h} +  \bar{\bm{\mu}} \cdot [(\bm{k}\cdot\bm{u})^{-1}\bm{N}  +  \tensor[^e]{\bm{\gamma}}{}] \cdot \bar{\bm{\varepsilon}}\cdot[ (\bm{k}\cdot\bm{u})^{-1}\bm{M} + \tensor[^h]{\bm{\gamma}}{} ]
\end{equation}
and where $\bar{\bm{\varepsilon}}$ is the restricted inverse of $\bm{\varepsilon}$, such that $\bar{\bm{\varepsilon}}\cdot\bm{\varepsilon} = \bm{\varepsilon}\cdot\bar{\bm{\varepsilon}} = \bm{h}$.
Note that since $\bm{\varepsilon}$ is transverse, then $\bm{\varepsilon} = \bm{h}\cdot\bm{\varepsilon}\cdot\bm{h}$, which means that 
\begin{equation}
 \mathrm{adj}(\bm{\varepsilon}) = \mathrm{adj}(\bm{h}\cdot\bm{\varepsilon}\cdot\bm{h}) = \mathrm{adj}(\bm{h})\cdot\mathrm{adj}(\bm{\varepsilon})\cdot\mathrm{adj}(\bm{h}).
\end{equation}
But since
\begin{equation} \label{Eq:Adjh}
 \mathrm{adj}(\bm{h}) = (\bm{u}^{\flat}\cdot \bm{u})^{-1} \bm{u}^{\flat}\otimes\bm{u},
\end{equation}
then
\begin{equation}
  \mathrm{adj}(\bm{\varepsilon}) = (\bm{u}^{\flat}\cdot \bm{u})^{-2} (\bm{u}\cdot\mathrm{adj}(\bm{\varepsilon})\cdot\bm{u}^{\flat}) \bm{u}^{\flat}\otimes\bm{u}.
\end{equation}
This brings us to the final form
\begin{equation}
 \mathrm{adj}(\bm{X}) = P (\bm{k}\otimes \bm{k}^{\sharp})
\end{equation}
as postulated, where 
\begin{equation} \label{Eq:P}
 P = -\frac18(\bm{u}^{\flat}\cdot\bm{u})^{-4} (\bm{k}\cdot\bm{u})^4( \bm{u}.\mathrm{adj}\left( \bm{\varepsilon} \right)\cdot\bm{u}^{\flat})( \bm{u}\cdot\mathrm{adj}\left( \bm{Q} \right)\cdot\bm{u}^{\flat}).
\end{equation}
Thus, I have shown that the requirement $\det(X)=0$, which is satisfied identically for wave solutions of Maxwell's equations, provides the scalar condition $P=0$ with $P$ given above.
This subsection began with the form of $\mc$ given in terms of the traditional parameters by Eq.~(\ref{Eq:ChiTradParameters}).  
This was not necessary, and one could easily reexpress $\bm{Q}$ in terms of the original set of transverse constitutive parameters derived from $\bm{G} = \star\mc\bm{F}$, but the approach followed provides a connection to the set of parameters that are typically more familiar, and remarkably, they facilitate the factorization of $P$ to determine the optical metric and hence the light cones of the medium, the subject of the next section.

\section{The Optical Metric} \label{Sec:OpticalMetric}
The concept of the optical metric has a long history.  From the early days of general relativity it was known that the light-deflecting properties of a massive spherical object could be obtained without general relativity if the space around the object were filled with an appropriately refracting medium \cite{Eddington1920}.
Gordon reversed this observation by asking whether a refracting medium could be mathematically represented as a curved vacuum space-time, and found the optical metric for isotropic dielectric media moving in a background space-time \cite{Gordon1923}.  
Tamm studied anisotropic media residing in a Minkowski background space-time, but did not find the optical metric for such media \cite{Tamm1924}.
Plebanski identified a mapping from a general curved space-time into a dielectric media residing in a flat Minkowski space-time \cite{Plebanski1960pr}, which essentially provides an avenue to map a medium into an optical metric.
However, since Plebanski's derivation starts from the vacuum it is restricted to nonbirefringent media.
The approach by Balakin and Zimdahl postulates two optical metrics and then seeks to match them with the material parameters \cite{Balakin2005gerg}, but does not appear to consider the background space-time in which the medium resides.
The idea of the optical metric has received a great deal of attention over the years from different perspectives \cite{Hehl2002ijmpa,Lammerzahl2004prd,Hehl,Itin2009jpa,Schuster2017prd}, many of which have a focus on the conditions under which the optical metric becomes pseudo-Riemannian.
By contrast, the derivation here seeks a tractable expression for the optical metric in terms of the familiar transverse parameters; beginning with Maxwell's equations inside a general medium residing in a background space-time, and showing how the optical metric emerges in the geometrical optics limit.

As previously mentioned, $P$ is a fourth order polynomial in $\bm{k}$, and wave solutions of Maxwell's equations must satisfy $P=0$.
For normalized observers in vacuum, $P = [\frac12\bm{g}^{-1}(\bm{k}\otimes\bm{k})]^2 = [\frac12 g^{\mu\nu}k_{\mu}k_{\nu}]^2=0$, which shows a degeneracy in the solutions reflecting the fact that the vacuum is not birefringent and all polarization states see the same light cone.
The strategy for media in a background space-time will be to show that $P$ can be written in the form of two pseudo-Finslerian optical metrics
\begin{equation}
 P \propto \left[\frac12 \bm{\frak{g}}^{-1}_+(\bm{k}\otimes\bm{k})\right]\left[\frac12 \bm{\frak{g}}^{-1}_-(\bm{k}\otimes\bm{k})\right],
\end{equation}
that are degenerate for certain types of media and reduce to $\bm{g}^{-1}$ in vacuum.

The remaining unknown quantity in $P$ is $\bm{u}\cdot\mathrm{adj}\left( \bm{Q} \right)\cdot\bm{u}^{\flat}$, a tensorial, index expression for which is 
\begin{equation} \label{Eq:AdjQ}
 \bm{u}\cdot\mathrm{adj}\left( \bm{Q} \right)\cdot\bm{u}^{\flat} = \frac16 u^{\alpha}u_{\beta} \delta_{\alpha\beta_1\beta_2\beta_3}^{\beta\alpha_1\alpha_2\alpha_3} Q\indices{_{\alpha_1}^{\beta_1}}
 Q\indices{_{\alpha_2}^{\beta_2}} Q\indices{_{\alpha_3}^{\beta_3}}.
\end{equation}
where $\delta_{\alpha\beta_1\beta_2\beta_3}^{\beta\alpha_1\alpha_2\alpha_3}$ is the generalized Kronecker delta.
Thus we need to consider the cube of the complicated matrix expression for $\bm{Q}$ in Eq.~(\ref{Eq:Q}), and then apply some complicated combinatorics to each term.
To do so with the most general possible $\mc$ can lead one in dizzying circles for months. Hence, to isolate the critical features of the optical metric I will make a simplifying assumption on the parameters but will try to keep them as general as possible.
It turns out that most of the complexity in the optical metric comes from the magnetoelectric terms $\tensor[^h]{\bm{\gamma}}{}$ and $\tensor[^e]{\bm{\gamma}}{}$.  To see this, consider the decomposition of $\tensor[^h]{\bm{\gamma}}{}$ and $\tensor[^e]{\bm{\gamma}}{}$ into antisymmetric, trace, and traceless-symmetric parts as
\begin{equation} \label{Eq:GammahDecomposition}
\tensor[^h]{\bm{\gamma}}{} = \accentset{h}{\bm{\gamma}}^{\sharp} \cdot\star(\bm{u}^{\flat}\wedge\bm{h}) +  Tr(\tensor[^h]{\bm{\gamma}}{})\bm{h} + \tensor[^h]{\bm{\gamma}}{^S},
\end{equation}
and
\begin{equation} \label{Eq:GammaeDecomposition}
 \tensor[^e]{\bm{\gamma}}{} = (\bm{h}\wedge\bm{u})\star \cdot \accentset{e}{\bm{\gamma}} +  Tr(\tensor[^e]{\bm{\gamma}}{})\bm{h} + \tensor[^e]{\bm{\gamma}}{^S}.
\end{equation}
Here, $\accentset{h}{\bm{\gamma}}$ and $\accentset{e}{\bm{\gamma}}$ are 1-forms, while $\tensor[^e]{\bm{\gamma}}{^S}$ and $\tensor[^h]{\bm{\gamma}}{^S}$ are traceless and symmetric with respect to the background metric, i.e. when both indices are either down or up.
By comparing with the definitions of $\bm{M}$ and $\bm{N}$ in Eqs.~(\ref{Eq:M}) and (\ref{Eq:N}), one can see that the antisymmetric parts of  $\tensor[^h]{\bm{\gamma}}{}$ and $\tensor[^e]{\bm{\gamma}}{}$ will naturally combine with $\bm{M}$ and $\bm{N}$ in a special way, while the trace and traceless-symmetric parts will contribute in a much more complicated manner.
Assume, therefore, that  $\tensor[^{e}]{\bm{\gamma}}{}$ and $\tensor[^{h}]{\bm{\gamma}}{}$ are each purely antisymmetric and that
\begin{equation}
  Tr(\tensor[^h]{\bm{\gamma}}{}) =  Tr(\tensor[^e]{\bm{\gamma}}{}) = \tensor[^e]{\bm{\gamma}}{^S} = \tensor[^h]{\bm{\gamma}}{^S} = 0.
\end{equation}
This assumption imposes 12 constraints on $\mc$, effectively reducing the number of free parameters from 36 to 24, which is still reasonably general, but significantly reduces the complexity involved in finding the optical metric.
One might think that this condition simply reduces the number of free parameters in $\bm{\chi}$ by setting some of them to zero, for example that all the magnetoelectric contributions from the skewon part of $\bm{\chi}$, $\tensor[^s]{\bm{\chi}}{}$ vanish.  
This is not quite the case because we are dealing with $\tensor[^{(h/e)}]{\bm{\gamma}}{}$ rather than $\tensor[^{(b/e)}]{\bm{\gamma}}{^c}$ which are related by $\bm{\mu}$, but thus far $\bm{\mu}$ has had no symmetry conditions imposed on it.

With the preceding assumption on the magnetoelectric couplings, the complexity is reduced because it allows $\bm{Q}$ to be written as  
\begin{equation}
  \bm{Q} = -\bm{h} +  (\bm{k}\cdot\bm{u})^{-2} \bar{\bm{\mu}} \cdot [(\bm{h}\wedge\bm{u})\star \cdot (\bm{k} +(\bm{k}\cdot\bm{u}) \accentset{e}{\bm{\gamma}} )] \cdot \bar{\bm{\varepsilon}}\cdot [ (\bm{k} +(\bm{k}\cdot\bm{u}) \accentset{h}{\bm{\gamma}})^{\sharp}\cdot \star(\bm{u}^{\flat}\wedge \bm{h} )].
\end{equation}
This can be rewritten with some more tractable notation. Let
\begin{equation} \label{Eq:Zform}
 \begin{aligned} 
  \accentset{e}{\bm{Z}} & = (\bm{\delta}\wedge\bm{u})\star \cdot (\bm{k} +(\bm{k}\cdot\bm{u}) \accentset{e}{\bm{\gamma}}), \\
  \accentset{h}{\bm{Z}} & = (\bm{\delta}\wedge\bm{u})\star \cdot (\bm{k} +(\bm{k}\cdot\bm{u}) \accentset{h}{\bm{\gamma}}),
 \end{aligned} 
\end{equation}
and
\begin{equation}
 \bm{W} = \bar{\bm{\mu}}\cdot  \accentset{e}{\bm{Z}} \cdot \bar{\bm{\varepsilon}} \cdot  \accentset{h}{\bm{Z}}.
\end{equation}
Then the tensor $\bm{Q}$ has the rather more appealing form
\begin{equation}
 \bm{Q} = -\bm{h} + (\bm{k}\cdot\bm{u})^{-2}\bm{W}
\end{equation}
from which it becomes much more straightforward to calculate the required adjugate.  Expanding Eq.~(\ref{Eq:AdjQ}) leads to
\begin{multline}
 \bm{u}\cdot\mathrm{adj}(\bm{Q})\cdot\bm{u}^{\flat} = -\frac16 u^{\alpha}u_{\beta}\delta_{\alpha\beta_1\beta_2\beta_3}^{\beta\alpha_1\alpha_2\alpha_3} h_{\alpha_1}^{\beta_1}h_{\alpha_2}^{\beta_2}h_{\alpha_3}^{\beta_3} + \frac36 (\bm{k}\cdot\bm{u})^{-2} u^{\alpha}u_{\beta} \delta_{\alpha\beta_1\beta_2\beta_3}^{\beta\alpha_1\alpha_2\alpha_3} h_{\alpha_1}^{\beta_1}h_{\alpha_2}^{\beta_2} (\bm{W})\indices{_{\alpha_3}^{\beta_3}} \\ -  \frac36 (\bm{k}\cdot\bm{u})^{-4} u^{\alpha}u_{\beta} \delta_{\alpha\beta_1\beta_2\beta_3}^{\beta\alpha_1\alpha_2\alpha_3} h_{\alpha_1}^{\beta_1}(\bm{W})\indices{_{\alpha_2}^{\beta_2}} (\bm{W})\indices{_{\alpha_3}^{\beta_3}}.
\end{multline}
The reason there is no $(\bm{k}\cdot\bm{u})^{-6}$ term is that
\begin{equation}
 u^{\alpha}u_{\beta} \delta_{\alpha\beta_1\beta_2\beta_3}^{\beta\alpha_1\alpha_2\alpha_3} (\accentset{h}{Z}\indices{_{\alpha_1}^{\beta_1}}) (\accentset{h}{Z}\indices{_{\alpha_2}^{\beta_2}}) (\accentset{h}{Z}\indices{_{\alpha_3}^{\beta_3}}) = (\bm{u}^{\flat}\cdot\bm{u}) h_{\beta_1\beta_2\beta_3}^{\alpha_1\alpha_2\alpha_3} (\accentset{h}{Z}\indices{_{\alpha_1}^{\beta_1}}) (\accentset{h}{Z}\indices{_{\alpha_2}^{\beta_2}}) (\accentset{h}{Z}\indices{_{\alpha_3}^{\beta_3}}) = 0.
\end{equation}
This is because the expression in the middle is proportional to the three-dimensional determinant of $\accentset{h}{\bm{Z}}$ in the transverse subspace, but it can be readily seen in Eq.~(\ref{Eq:Zform}) that $\accentset{h}{\bm{Z}}$ is annihilated by both $\bm{u}$ and $\bm{k} +(\bm{k}\cdot\bm{u}) \accentset{h}{\bm{\gamma}}$ and hence is actually only two dimensional, so its three-dimensional determinant vanishes identically.

Next, consider the action of the generalized Kronecker delta on products of $\bm{h}$.  One finds
\begin{subequations}
 \begin{equation}
  u^{\alpha}u_{\beta}\delta_{\alpha\beta_1\beta_2\beta_3}^{\beta\alpha_1\alpha_2\alpha_3} h_{\alpha_1}^{\beta_1}h_{\alpha_2}^{\beta_2}h_{\alpha_3}^{\beta_3} = 6(\bm{u}^{\flat}\cdot\bm{u})
 \end{equation}
 \begin{equation}
  u^{\alpha}u_{\beta} \delta_{\alpha\beta_1\beta_2\beta_3}^{\beta\alpha_1\alpha_2\alpha_3} h_{\alpha_1}^{\beta_1}h_{\alpha_2}^{\beta_2} = 2  (\bm{u}^{\flat}\cdot\bm{u}) h_{\beta_3}^{\alpha_3}
 \end{equation}
 \begin{equation}
  u^{\alpha}u_{\beta} \delta_{\alpha\beta_1\beta_2\beta_3}^{\beta\alpha_1\alpha_2\alpha_3} h_{\alpha_1}^{\beta_1} = (\bm{u}^{\flat}\cdot\bm{u}) h_{\beta_2\beta_3}^{\alpha_2\alpha_3}.
 \end{equation}
\end{subequations}
With this, we now have
\begin{equation}
 \bm{u}\cdot\mathrm{adj}(\bm{Q})\cdot\bm{u}^{\flat} = -(\bm{u}^{\flat}\cdot\bm{u})\left[1 - (\bm{k}\cdot\bm{u})^{-2} Tr(\bm{W})
 - \frac12 (\bm{k}\cdot\bm{u})^{-4} \left(Tr(\bm{W}\cdot\bm{W}) - Tr(\bm{W})^2 \right)\right].
\end{equation}
Completing the square, this can be factored to
\begin{multline}
 \bm{u}\cdot\mathrm{adj}(\bm{Q})\cdot\bm{u}^{\flat} =  -(\bm{u}^{\flat}\cdot\bm{u})\left[1 - \frac12 (\bm{k}\cdot\bm{u})^{-2} Tr(\bm{W}) 
 + (\bm{k}\cdot\bm{u})^{-2} \sqrt{ \frac12 Tr(\bm{W}\cdot\bm{W}) -\frac14 Tr(\bm{W})^2}\right] 
 \\ \times
 \left[1 - \frac12 (\bm{k}\cdot\bm{u})^{-2} Tr(\bm{W}) 
  - (\bm{k}\cdot\bm{u})^{-2} \sqrt{ \frac12 Tr(\bm{W}\cdot\bm{W}) -\frac14 Tr(\bm{W})^2}\right].
\end{multline}
Returning now to Eq.~(\ref{Eq:P}) for $P$, we have
\begin{equation} \label{Eq:Pfactored}
 P = \frac12 (\bm{u}^{\flat}\cdot\bm{u}) ( \bm{u}\cdot\mathrm{adj}\left( \bm{\varepsilon} \right)\cdot\bm{u}^{\flat})^{-1}H_+ H_-
\end{equation}
with
\begin{equation} \label{Eq:Hplus}
 H_+ = \frac12 (\bm{u}^{\flat}\cdot\bm{u})^{-2}( \bm{u}\cdot\mathrm{adj}\left( \bm{\varepsilon} \right)\cdot\bm{u}^{\flat}) \left[(\bm{k}\cdot\bm{u})^2- \frac12  Tr(\bm{W}) 
  +  \sqrt{ \frac12 Tr(\bm{W}\cdot\bm{W}) -\frac14 Tr(\bm{W})^2}\right] 
\end{equation}
and
\begin{equation} \label{Eq:Hminus}
 H_- = \frac12 (\bm{u}^{\flat}\cdot\bm{u})^{-2}( \bm{u}\cdot\mathrm{adj}\left( \bm{\varepsilon} \right)\cdot\bm{u}^{\flat}) \left[(\bm{k}\cdot\bm{u})^2- \frac12  Tr(\bm{W}) 
  -  \sqrt{ \frac12 Tr(\bm{W}\cdot\bm{W}) -\frac14 Tr(\bm{W})^2}\right]. 
\end{equation}

Recall that wave solutions of Maxwell's equations must satisfy $P=0$.  
This condition can in turn be met through the satisfaction of either of the conditions $H_+=0$ or $H_-=0$.  
In other words, the medium exhibits two wave-propagation eigenstates that follow different ray trajectories, i.e.\ birefringence.
The $H_{\pm}$ serve as pseudo-Finslerian structures on the manifold, and we may define the associated optical metrics
\begin{equation} \label{Eq:OpticalMetrics}
 \frak{g}_{\pm}^{\mu\nu}(\bm{x},\bm{k}) = \frac{\partial^2 H_{\pm}}{\partial k_{\mu} \partial k_{\nu}},
\end{equation}
where the arguments of $\bm{\frak{g}}_{\pm}^{-1}$ have been included explicitly to show that the optical metric depends not just on the location in the medium but also on the wave covector at any point. 
This dependence on the wave covector means that waves passing through the same point but propagating in different directions will in general see different optical metrics.
This complicated dependence of the optical metrics on the wave covector potentially makes them a somewhat less useful concept than one might otherwise expect.
The optical metric defines the light cone, but the residual dependence on $\bm{k}$ may prove to make an algebraic determination of the light cone in practice rather difficult.  
Instead, all of the ray content for each polarization is contained in $H_{\pm}$, as discussed in Sec.\ref{Sec:Kinematics}.

The expressions for $H_{\pm}$ given above are not particularly illuminating.  Ideally, we should be able to isolate the $\bm{k}$ dependence in an expression of the form
\begin{equation}
 H_{\pm} = a^{\alpha\beta}k_{\alpha}k_{\beta} \pm \sqrt{ b^{\mu\nu\sigma\rho}k_{\mu}k_{\nu}k_{\sigma}k_{\rho}}
\end{equation}
where $a^{\alpha\beta}$ and $b^{\mu\nu\sigma\rho}$ depend in some way on the usual material parameters as functions of points on the manifold, but are independent of $\bm{k}$.  Examining $\bm{W}$, one finds
\begin{equation} \label{Eq:Windices}
\begin{aligned} 
 \bm{W} & = \bar{\bm{\mu}} \cdot [(\bm{\delta}\wedge\bm{u})\star \cdot (\bm{k} + (\bm{k}\cdot\bm{u}) \accentset{e}{\bm{\gamma}})] \cdot \bar{\bm{\varepsilon}}\cdot  [ (\bm{k} +(\bm{k}\cdot\bm{u}) \accentset{h}{\bm{\gamma}})^{\sharp}\cdot \star(\bm{u}^{\flat}\wedge \bm{\delta} )] \\
 W\indices{_{\alpha}^{\kappa}}& = -(\bm{u}^{\flat}\cdot\bm{u}) h^{\lambda\kappa\rho}_{\psi\beta\varphi} g_{\lambda\tau}\bar{\varepsilon}\indices{_{\sigma}^{\tau}}g^{\sigma\psi} \bar{\mu}\indices{_{\alpha}^{\beta}} g^{\eta\varphi} (\delta_{\rho}^{\mu} + \accentset{e}{\bm{\gamma}}_{\rho}u^{\mu})  (\delta_{\eta}^{\nu} + \accentset{e}{\bm{\gamma}}_{\eta}u^{\nu}) k_{\mu}k_{\nu} \\
 & = W\indices{_{\alpha}^{\kappa\mu\nu}}k_{\mu}k_{\nu}
\end{aligned} 
\end{equation}
from which it follows that $H_{\pm}$ attains the desired format
\begin{equation} \label{Eq:Hpm}
 H_{\pm} = \frac12 (\bm{u}^{\flat}\cdot\bm{u})^{-2}( \bm{u}.\mathrm{adj}\left( \bm{\varepsilon} \right).\bm{u}^{\flat}) 
   \left[( u^{\mu}u^{\nu} - \frac12 W\indices{_{\alpha}^{\alpha\mu\nu}})k_{\mu}k_{\nu} \pm  \sqrt{ \left(\frac12 W\indices{_{\alpha}^{\beta\mu\nu}}W\indices{_{\beta}^{\alpha\sigma\rho}} -\frac14 W\indices{_{\alpha}^{\alpha\mu\nu}} W\indices{_{\beta}^{\beta\sigma\rho}} \right) k_{\mu}k_{\nu}k_{\sigma}k_{\rho} } \right]. 
\end{equation}

With this explicit form of the dependence on $\bm{k}$, it is straightforward to show that the optical metrics may be computed as in Eq.~(\ref{Eq:OpticalMetrics}), and one may subsequently show that
\begin{equation} \label{Eq:H_as_optical_metric}
 H_{\pm} = \frac12 \frak{g}_{\pm}^{\alpha\beta}k_{\alpha}k_{\beta} = 0
\end{equation}
provides the condition satisfied by light rays in dielectric media.
Although the index expression Eq.~(\ref{Eq:Windices}) for $W\indices{_{\alpha}^{\pi\mu\nu}}$ looks complicated, it is entirely expressed in terms of the ``usual'' transverse parameters $\bm{\varepsilon}$, $\bm{\mu}$, $\tensor[^h]{\bm{\gamma}}{}$, and $\tensor[^e]{\bm{\gamma}}{}$, and the background metric, and it may be easily handled by any computer algebra system.

\section{Reduction to pseudo-Riemannian optical metric} \label{Sec:RiemannReduction}
A natural question of interest is, under what conditions does birefringence vanish?  
For nonbirefringent media, the independent light cones for each polarization become degenerate and $\bm{\mathfrak{g}}_+=\bm{\mathfrak{g}}_-$, or equivalently, $H_+=H_-$. 
This condition requires the square root term of Eqs.~(\ref{Eq:Hplus}) and (\ref{Eq:Hminus}), or equivalently Eq.~(\ref{Eq:Hpm}), to vanish, whence the two associated pseudo-Finslerian optical metrics defined by Eq.~(\ref{Eq:OpticalMetrics}) degenerate to a single, pseudo-Riemannian optical metric. 
For any given medium there may exist particular choices of $\bm{k}$ for which the square root term vanishes, corresponding to optical axes of the medium.
More generally, one would like to know which constraints on the medium must be imposed such that the square root vanishes for \textit{all} $\bm{k}$, which is a more difficult question to answer.
There are several possibilities that should be investigated, but from the complexity of the expression it seems clear that it will only vanish under quite restrictive conditions.
In particular, consider the restrictions $\accentset{e}{\bm{\gamma}} = \accentset{h}{\bm{\gamma}} = \bm{\gamma}$, and $\varepsilon_{\alpha\beta}$ and $\xi_{\alpha\beta}$ symmetric with respect to the background metric.

First, return to Eq.~(\ref{Eq:Windices}) for $\bm{W}$, and let
\begin{equation}
 \bm{q} = \bm{k} + (\bm{k}\cdot\bm{u}) \bm{\gamma};
\end{equation}
then $\bm{W}$ has the simplified expression
\begin{equation} \label{Eq:WSimpleIndex}
 W\indices{_{\alpha}^{\kappa}} = -(\bm{u}^{\flat}\cdot\bm{u}) (h^{\lambda\kappa\rho}_{\psi\beta\varphi}q_{\rho}q^{\varphi}) \bar{\varepsilon}\indices{_{\lambda}^{\psi}} \bar{\mu}\indices{_{\alpha}^{\beta}}
\end{equation}
(the metric has been absorbed by the symmetry of $\bm{\varepsilon}$).
Notice that now $W\indices{_{\alpha}^{\kappa}}$ is actually orthogonal to \textit{both} $\bm{u}$ and $\bm{q}$, and is therefore effectively only two-dimensional.
Define
\begin{equation}
 \bm{j} = \bm{h} - \frac{(\bm{h}\cdot\bm{q})\otimes (\bm{h}\cdot\bm{q})^{\sharp}}{(\bm{h}\cdot\bm{q})\cdot (\bm{h}\cdot\bm{q})^{\sharp}}
\end{equation}
as the projection operator from the subspace orthogonal to $\bm{u}$ to the subspace orthogonal to both $\bm{u}$ and $\bm{q}$.  One may readily show that $\bm{j}\cdot\bm{j}=\bm{j}$ and that  the Kronecker $\bm{h}$ on the subspace orthogonal to $\bm{u}$, is projected to $\bm{j}\cdot\bm{h}\cdot\bm{j} = \bm{j}$, and thus $\bm{j}$ serves as the Kronecker tensor on the subspace orthogonal to both $\bm{u}$ and $\bm{q}$.
Similar to Eq.~(\ref{Eq:hGeneralized}), one finds
\begin{equation}
 h^{\sigma\kappa\rho}_{\tau\theta\varphi} q_{\rho}q^{\varphi} = \left((\bm{h}\cdot\bm{q})\cdot (\bm{h}\cdot\bm{q})^{\sharp}\right) j^{\sigma\kappa}_{\tau\theta}.
\end{equation}
Equation (\ref{Eq:WSimpleIndex}) for $W\indices{_{\alpha}^{\kappa}}$ now becomes
\begin{equation} 
 W\indices{_{\alpha}^{\kappa}} = -(\bm{u}^{\flat}\cdot\bm{u}) \left((\bm{h}\cdot\bm{q})\cdot (\bm{h}\cdot\bm{q})^{\sharp}\right) j^{\lambda\kappa}_{\psi\beta} \bar{\varepsilon}\indices{_{\lambda}^{\psi}} \bar{\mu}\indices{_{\alpha}^{\beta}} .
\end{equation}
Calculating the argument of the square root in $H_{\pm}$ in Eq.~(\ref{Eq:Hpm}), one finds
\begin{equation}
  \frac12 Tr(\bm{W}\cdot\bm{W}) -\frac14 Tr(\bm{W})^2 =  (\bm{u}^{\flat}\cdot\bm{u})^2  \left((\bm{h}\cdot\bm{q})\cdot (\bm{h}\cdot\bm{q})^{\sharp}\right)^2
  \left[\frac12 j^{\sigma\kappa}_{\tau\theta} j^{\mu\rho}_{\nu\beta}  \bar{\varepsilon}\indices{_{\sigma}^{\tau}}\bar{\varepsilon}\indices{_{\mu}^{\nu}} \bar{\mu}\indices{_{\kappa}^{\beta}}\bar{\mu}\indices{_{\rho}^{\theta}} 
   - \frac14 (j^{\sigma\kappa}_{\tau\theta}  \bar{\varepsilon}\indices{_{\sigma}^{\tau}} \bar{\mu}\indices{_{\kappa}^{\theta}})^2 \right].
\end{equation}
Expanding the generalized Kronecker tensors into products of $\bm{j}$, this becomes
\begin{multline}  \label{Eq:SymmetricRoot}
  \frac12 Tr(\bm{W}\cdot\bm{W}) -\frac14 Tr(\bm{W})^2 = (\bm{u}^{\flat}\cdot\bm{u})^2  \left((\bm{h}\cdot\bm{q})\cdot (\bm{h}\cdot\bm{q})^{\sharp}\right)^2 
  \left[\frac12 Tr(\bar{\bm{\varepsilon}}_{\bm{j}})^2 Tr(\bar{\bm{\mu}}_{\bm{j}}^2)
  - Tr(\bar{\bm{\varepsilon}}_{\bm{j}})Tr(\bar{\bm{\varepsilon}}_{\bm{j}}\cdot\bar{\bm{\mu}}_{\bm{j}}^2 )
  +\frac12 Tr(\bar{\bm{\varepsilon}}_{\bm{j}}^2\cdot\bar{\bm{\mu}}_{\bm{j}}^2 ) 
  \right. \\ \left.  -\frac14 Tr(\bar{\bm{\varepsilon}}_{\bm{j}})^2Tr(\bar{\bm{\mu}}_{\bm{j}})^2
  + \frac12 Tr(\bar{\bm{\varepsilon}}_{\bm{j}})Tr(\bar{\bm{\mu}}_{\bm{j}}) Tr(\bar{\bm{\varepsilon}}_{\bm{j}}\cdot\bar{\bm{\mu}}_{\bm{j}} )
  - \frac14  Tr(\bar{\bm{\varepsilon}}_{\bm{j}}\cdot\bar{\bm{\mu}}_{\bm{j}} )^2
  \right]
\end{multline}
where
\begin{equation}
 \bar{\bm{\varepsilon}}_{\bm{j}} = \bm{j}\cdot\bar{\bm{\varepsilon}}\cdot\bm{j}, \quad \bar{\bm{\mu}}_{\bm{j}} = \bm{j}\cdot\bar{\bm{\mu}}\cdot\bm{j}.
\end{equation}
This will vanish for two possible cases.
\subsubsection{$\bar{\bm{\varepsilon}}$ and $\bar{\bm{\mu}}$ proportional to $\bm{h}$}
Consider first the case where $\bar{\bm{\varepsilon}}$ and $\bar{\bm{\mu}}$ are both proportional to $\bm{h}$; i.e., they are both isotropic so $\bm{\varepsilon}=\varepsilon\bm{h}$ and $\bm{\mu}=\mu\bm{h}$.  
Since $\bm{j}\cdot\bm{h}\cdot\bm{j}=\bm{j}$, the sum over traces in Eq.~(\ref{Eq:SymmetricRoot}) becomes
\begin{equation}
\begin{aligned}
  & \varepsilon^{-2}\mu^{-2}  \left[\frac12 Tr(\bm{j})^2 Tr(\bm{j}^2)
   - Tr(\bm{j})Tr(\bm{j}\cdot\bm{j}^2)
   +\frac12 Tr(\bm{j}^2\cdot\bm{j}^2 ) 
    -\frac14 Tr(\bm{j})^2 Tr(\bm{j})^2
   + \frac12 Tr(\bm{j})Tr(\bm{j}) Tr(\bm{j}\cdot\bm{j})
   - \frac14  Tr(\bm{j}\cdot\bm{j})^2
   \right]
 \\ & = \varepsilon^{-2}\mu^{-2} \left[\frac12 Tr(\bm{j})^2 Tr(\bm{j})
    - Tr(\bm{j})Tr(\bm{j})
    +\frac12 Tr(\bm{j} ) 
     -\frac14 Tr(\bm{j})^2 Tr(\bm{j})^2
    + \frac12 Tr(\bm{j})^2 Tr(\bm{j})
    - \frac14  Tr(\bm{j})^2
    \right]
 \\ & = 0
\end{aligned}
\end{equation}
where the final equality follows from the fact that $Tr(\bm{j}) = 2$.

\subsubsection{$\bar{\bm{\mu}} = \bar{\bm{\varepsilon}}$ (equivalently $\bm{\mu} = \bm{\varepsilon}$)}
Next, consider the case in which $\bm{\mu}=\bm{\varepsilon}$.  In this case the sum of traces becomes
\begin{equation}
 \frac14\left[ -Tr(\bar{\bm{\mu}}_{\bm{j}})^4 + 4Tr(\bar{\bm{\mu}}_{\bm{j}})^2Tr(\bar{\bm{\mu}}_{\bm{j}}^2) - 4Tr(\bar{\bm{\mu}}_{\bm{j}})Tr(\bar{\bm{\mu}}_{\bm{j}}^3) - Tr(\bar{\bm{\mu}}_{\bm{j}}^2)^2 + 2Tr(\bar{\bm{\mu}}_{\bm{j}}^4) \right].
\end{equation}
To see that this vanishes, rewrite it in the following clever way
\begin{multline} \label{Eq:CleverRewrite}
 -\frac{2}{24}\left[ Tr(\bar{\bm{\mu}}_{\bm{j}})^4 - 6Tr(\bar{\bm{\mu}}_{\bm{j}})^2 Tr(\bar{\bm{\mu}}_{\bm{j}}^2) + 3 Tr(\bar{\bm{\mu}}_{\bm{j}}^2)^2 + 8 Tr(\bar{\bm{\mu}}_{\bm{j}})Tr(\bar{\bm{\mu}}_{\bm{j}}^3) -6 Tr(\bar{\bm{\mu}}_{\bm{j}}^4)\right] 
 \\ - \frac16 Tr(\bar{\bm{\mu}}_{\bm{j}}) \left[ Tr(\bar{\bm{\mu}}_{\bm{j}})^3 - 3Tr(\bar{\bm{\mu}}_{\bm{j}})Tr(\bar{\bm{\mu}}_{\bm{j}}^2) +2 Tr(\bar{\bm{\mu}}_{\bm{j}}^3)\right].
\end{multline}
Now, each of the bracketed terms is the trace expression for the determinant of a matrix, the first being the four-dimensional determinant and the second being the three-dimensional determinant, in other words, Eq.~(\ref{Eq:CleverRewrite}) is equivalent to
\begin{equation}
 -2 Det_4(\bar{\bm{\mu}}_{\bm{j}}) - Tr(\bar{\bm{\mu}}_{\bm{j}}) Det_3(\bar{\bm{\mu}}_{\bm{j}}).
\end{equation} 
This vanishes because $\bar{\bm{\mu}}_{\bm{j}}$, being the projection of $\bar{\bm{\mu}}$ into the two-dimensional subspace orthogonal to both $\bm{u}$ and $\bm{q}$, is effectively only two dimensional, and hence has identically vanishing three- and four-dimensional determinants.  We may conclude that a medium with $\bar{\bm{\mu}} = \bar{\bm{\varepsilon}}$ and  $\accentset{e}{\bm{\gamma}} = \accentset{h}{\bm{\gamma}}$ is also not birefringent, and the pseudo-Finslerian optical metric becomes pseudo-Riemannian in these two cases.

\subsection{Expression for pseudo-Riemannian optical metric}
Now that the conditions under which $H_{\pm}$ admit a pseudo-Riemannian optical metric have been identified we may find a more explicit expression for it in terms of the usual material parameters.
Consider first the conditions adopted in Case 2 above; that $\bm{\varepsilon}$ and $\bm{\mu}$ are equal and symmetric with respect to the background space-time metric, and that $\tensor[^e]{\bm{\gamma}}{}$ and $\tensor[^h]{\bm{\gamma}}{}$ are antisymmetric with respect to the background space-time metric and defined as described in Eqs.~(\ref{Eq:GammahDecomposition}) and (\ref{Eq:GammaeDecomposition}) with $\accentset{e}{\bm{\gamma}} = \accentset{h}{\bm{\gamma}} =\bm{\gamma}$
Returning to the expression for $\bm{W}$ given by Eq.~(\ref{Eq:WSimpleIndex}) and setting $\bar{\bm{\varepsilon}} = \bar{\bm{\mu}}$,
\begin{equation}
  Tr(\bm{W}) = W\indices{_{\alpha}^{\alpha}} = -(\bm{u}^{\flat}\cdot\bm{u}) h^{\lambda\alpha\rho}_{\psi\beta\varphi} \bar{\mu}\indices{_{\lambda}^{\psi}} \bar{\mu}\indices{_{\alpha}^{\beta}}  q_{\rho}q^{\varphi}.
\end{equation}
From Eq.~(\ref{Eq:3DInverse}) for the three-dimensional inverse on the transverse subspace, it may be seen that 
\begin{equation}
 (\bm{u}^{\flat}\cdot\bm{u}) h^{\lambda\alpha\rho}_{\psi\beta\varphi} \bar{\mu}\indices{_{\lambda}^{\psi}} \bar{\mu}\indices{_{\alpha}^{\beta}} = 2(\bm{u}\cdot \textrm{adj}(\bar{\bm{\mu}})\cdot \bm{u}^{\flat}) \mu\indices{_{\varphi}^{\rho}}
\end{equation}
and thus
\begin{equation}
\begin{aligned} 
 Tr(\bm{W}) & = -2  (\bm{u}\cdot \textrm{adj}(\bar{\bm{\mu}})\cdot \bm{u}^{\flat}) \mu\indices{_{\psi}^{\theta}} g^{\psi\rho} q_{\rho}q_{\theta} \\
 & = -2(\bm{u}\cdot \textrm{adj}(\bar{\bm{\mu}})\cdot \bm{u}^{\flat}) \left(g^{\psi \rho} \mu\indices{_{\psi}^{\theta}}k_{\rho}k_{\theta} + g^{\psi \rho} \mu\indices{_{\psi}^{\nu}}\gamma_{\nu} k_\rho u^{\theta}k_{\theta}  + u^{\rho}k_{\rho}  \gamma_{\mu} g^{\psi \mu} \mu\indices{_{\psi}^{\theta}} k_{\theta} + \gamma_{\mu} g^{\psi \mu} \mu\indices{_{\psi}^{\nu}}\gamma_{\nu} u^{\rho}u^{\theta} k_{\rho}k_{\theta} \right) \\ 
 & = -2  (\bm{u}\cdot \textrm{adj}(\bar{\bm{\mu}})\cdot \bm{u}^{\flat}) \left(g^{\psi \rho} \mu\indices{_{\psi}^{\theta}} + g^{\psi \rho} \mu\indices{_{\psi}^{\nu}}\gamma_{\nu}  u^{\theta}  + u^{\rho}  \gamma_{\mu} g^{\psi \mu} \mu\indices{_{\psi}^{\theta}}  + \gamma_{\mu} g^{\psi \mu} \mu\indices{_{\psi}^{\nu}}\gamma_{\nu} u^{\rho}u^{\theta}  \right)k_{\rho}k_{\theta}.
\end{aligned} 
\end{equation}
Writing it in an index-free form
\begin{equation}
 Tr(\bm{W})= -2  (\bm{u}\cdot \textrm{adj}(\bar{\bm{\mu}})\cdot \bm{u}^{\flat}) \left[\bm{g}^{-1}\cdot \bm{\mu} + (\bm{g}^{-1}\cdot \bm{\mu} \cdot \bm{\gamma}) \otimes \bm{u} + \bm{u} \otimes (\bm{\gamma}\cdot \bm{g}^{-1}\cdot \bm{\mu}) +(\bm{\gamma}\cdot \bm{g}^{-1}\cdot \bm{\mu}\cdot \bm{\gamma}) \bm{u}\otimes\bm{u} \right](\bm{k}\otimes\bm{k}).
\end{equation}
Finally, returning to Eqs.~(\ref{Eq:Hplus}) and (\ref{Eq:Hminus}) and using the fact that
\begin{equation}
 (\bm{u}^{\flat}\cdot\bm{u})^{-2}( \bm{u}\cdot\mathrm{adj}\left( \bm{\mu} \right)\cdot\bm{u}^{\flat}) (\bm{u}\cdot \textrm{adj}(\bar{\bm{\mu}})\cdot \bm{u}^{\flat})  = 1,
\end{equation}
one has
\begin{multline}
 H_{\pm} = \frac12 \left[\left((\bm{u}^{\flat}\cdot\bm{u})^{-2}( \bm{u}\cdot\mathrm{adj}\left( \bm{\mu} \right)\cdot\bm{u}^{\flat})+(\bm{\gamma}\cdot \bm{g}^{-1}\cdot \bm{\mu}\cdot \bm{\gamma})\right) \bm{u}\otimes\bm{u} + (\bm{g}^{-1}\cdot \bm{\mu} \cdot \bm{\gamma}) \otimes \bm{u} + \bm{u} \otimes (\bm{\gamma}\cdot \bm{g}^{-1}\cdot \bm{\mu}) + \bm{g}^{-1}\cdot \bm{\mu}   \right](\bm{k}\otimes\bm{k}).
\end{multline}
Comparing with Eq.~(\ref{Eq:H_as_optical_metric}), it is clear that the optical metrics of the medium are degenerate, $\bm{\frak{g}}_+^{-1}=\bm{\frak{g}}_-^{-1} = \bm{\frak{g}}^{-1}$, and
\begin{equation} \label{Eq:OpticalMetric}
 \bm{\frak{g}}^{-1} = \left((\bm{u}^{\flat}\cdot\bm{u})^{-2}( \bm{u}\cdot\mathrm{adj}\left( \bm{\mu} \right)\cdot\bm{u}^{\flat})+(\bm{\gamma}\cdot \bm{g}^{-1}\cdot \bm{\mu}\cdot \bm{\gamma})\right) \bm{u}\otimes\bm{u} + (\bm{g}^{-1}\cdot \bm{\mu} \cdot \bm{\gamma}) \otimes \bm{u} + \bm{u} \otimes (\bm{\gamma}\cdot \bm{g}^{-1}\cdot \bm{\mu}) + \bm{g}^{-1}\cdot \bm{\mu}. 
\end{equation}

Although it required a rather lengthy derivation to obtain this expression, it may be verified by making a comparison between $P$ computed with Eq.~(\ref{Eq:Pfactored}) using this optical metric and $P$ computed directly from the adjugate of $\bm{X}$ as in Eq.~(\ref{Eq:P_as_adjX}).
Relative to the observer $\bm{u}$, the purely spatial part of the inverse metric is just $\bm{\mu}$ ($=\bm{\varepsilon}$) with its first index raised by the background space-time metric.
The time-space components of $\bm{\frak{g}}^{-1}$ are somewhat more complicated, but simplify when one considers $\bm{\frak{g}}$, where they become simply proportional to the magnetoelectric coupling 1-form $\bm{\gamma}$.  

It is straightforward to check that the optical metric reduces to the background space-time metric in vacuum. 
Indeed, setting $\bm{\mu}=\bm{h}$ and $\bm{\gamma} = 0$, and making use of Eq.~(\ref{Eq:Adjh}) for the adjugate of $\bm{h}$, one has
\begin{equation}
\begin{aligned} 
 \bm{\frak{g}}^{-1} & = (\bm{u}^{\flat}\cdot\bm{u})^{-1} \bm{u}\otimes\bm{u} + \bm{g}^{-1}\cdot\bm{h} \\
  & =  \bm{g}^{-1}\cdot\left(\frac{\bm{u}^{\flat}\otimes\bm{u}}{(\bm{u}^{\flat}\cdot\bm{u})} + \bm{h}\right)  = \bm{g}^{-1}\cdot\bm{\delta}  = \bm{g}^{-1}.  
\end{aligned}
\end{equation}

It is also straightforward to check that Gordon's optical metric is recovered for the case of isotropic, nonmagentoelectrically coupled media comoving with the observer \cite{Gordon1923}.
Indeed, setting $\bm{\varepsilon}=\varepsilon\bm{h}$, $\bm{\mu}=\mu\bm{h}$, and $\bm{\gamma} = 0$, and calculating $H_{\pm}$, one finds
\begin{equation}
  H_{\pm} = \frac12\left[\varepsilon (\bm{u}^{\flat}\cdot\bm{u})^{-1} \bm{u}\otimes\bm{u}  + \mu^{-1}\bm{g}^{-1}\cdot\bm{h} \right](\bm{k}\otimes\bm{k}).
\end{equation} 
Exploiting the conformal invariance of the metric we may rescale by $\mu$ and expand $\bm{h}$ to obtain
 \begin{equation}
  \bm{\frak{g}}^{-1} = \bm{g}^{-1} + (\bm{u}^{\flat}\cdot\bm{u})^{-1}(\varepsilon\mu -1 ) \bm{u}\otimes\bm{u}.
 \end{equation}
Setting the normalization to $(\bm{u}^{\flat}\cdot\bm{u})=-1$, the index expression
\begin{equation}
 \frak{g}^{\alpha\beta} = g^{\alpha\beta} - (\varepsilon\mu -1)u^{\alpha}u^{\beta}
\end{equation}
is identical to the optical metric found by Gordon \cite[Eq.~(16)]{Gordon1923}.

\section{Ray tracing and kinematics} \label{Sec:Kinematics}
To put the previous sections into a useful context and tie up the discussion of geometric optics, I turn now briefly to the question of ray tracing and the kinematics of congruences.  
This topic was recently covered in some detail in Ref.~\cite{Thompson2017prd}, where we studied the kinematics of light beams traversing dielectric media in curved background space-times, and derived the generalized Raychaudhuri equation associated with them.  
The Raychaudhuri equation describes how the cross section of a beam evolves along the length of the beam, and can essentially be thought of as describing the focus of the beam.
In Ref.~\cite{Thompson2017prd} we started with the proposition that $P$ is factorizable, deferring the proof for the present paper, and that the associated optical metrics exist in the form derived here.

In the geometric optics limit, solution data consist of space-time points $p$ together with a wave-(co)vector $\bm{k}$ at $p$.  In other words, solutions consist of curves in the eight-dimensional cotangent bundle $\mathcal{C}:\mathbb{R}\to T^*M$, i.e., parametrized curves in phase space 
\begin{equation}
 \mathcal{C}(\tau) = (p(\tau),\bm{k}(\tau)) = (x^{\mu}(\tau),k_{\mu}(\tau)).
\end{equation}
Thus, the geometric optics limit discards the tensorial nature of the fields and mathematically reduces the wave equation to the propagation of scalar particles.
A physical ray trajectory $\tilde{\mathcal{C}}$ is the projection of a solution curve $\mathcal{C}$ from the cotangent bundle into the manifold $M$ with the projection operator
\begin{equation}
 \pi: T^*M\to M
\end{equation}
such that
\begin{equation}
 \tilde{\mathcal{C}}(\tau) = \pi(\mathcal{C}(\tau)) = p(\tau).
\end{equation}
Obtaining these physical trajectories on the manifold is the goal of ray tracing, but it should be borne in mind that one must actually solve for the solution curve $\mathcal{C}$ in the cotangent bundle, and then project this to the manifold.

In the previous sections it was shown that solution curves must satisfy the condition $P=0$, and, at least under certain conditions on the magnetoelectric coupling, that $P\propto H_+H_-$.  
Thus any wave solution of Maxwell's equations in the geometrical optics limit must everywhere satisfy $H=0$ for either $H=H_+$ or $H=H_-$.
This means that $H=0$ anywhere along a solution curve $\mathcal{C}$ parametrized by $\tau$, and thus
\begin{equation}
 \frac{dH}{d\tau} =0.
\end{equation}
Since $H(p,\bm{k})$ is a function on the cotangent bundle, then
\begin{equation}
 \frac{dH}{d\tau} = \left(\frac{dx^{\mu}}{d\tau}\right)\frac{\partial H}{\partial x^{\mu}} + \left(\frac{dk_{\mu}}{d\tau}\right)\frac{\partial H}{\partial k_{\mu}} = \dot{x}^{\mu} \frac{\partial H}{\partial x^{\mu}} + \dot{k}_{\mu} \frac{\partial H}{\partial k_{\mu}} =0,
\end{equation}
which implies Hamilton's canonical equations
\begin{equation}
 \dot{x}^{\mu} = \frac{\partial H}{\partial k_{\mu}}, \quad \dot{k}_{\mu} = -\frac{\partial H}{\partial x^{\mu}}.
\end{equation}
The solution to this set of coupled equations determines the trajectories in the phase space, and the $x^{\mu}(\tau)$ component gives the ray trajectories on the manifold.
With $H$ as found in the previous sections, Hamilton's equations allow for ray tracing through dielectric media in curved space-times, where the medium may smoothly vary among positively refracting, negatively refracting, and vacuum.
Since the frequency component is traced over together with the spatial components of $\bm{k}$, they should automatically account for gravitational redshift and for frequency-shifting media \cite{Cummer2011jo}.

Let
\begin{equation}
 v^{\mu} = \dot{x}^{\mu}
\end{equation}
denote the tangent to $\tilde{\mathcal{C}}$.
Since $H=\frac12 \bm{\frak{g}}^{-1}(\bm{k}\otimes\bm{k})$ by Eq.~(\ref{Eq:H_as_optical_metric}), then by the first of Hamilton's equations it follows that
\begin{equation}
 \bm{v} = \bm{\frak{g}}^{-1}\cdot\bm{k}.
\end{equation}
Thus the ray direction is related to the wave vector through the optical metric.  In a vacuum space-time or in isotropic media one finds that $\bm{v}$ and $\bm{k}$ are ``parallel.''
The typical statement that $\bm{v}$ and $\bm{k}$ are \textit{not} parallel in more general media belies the fact that despite having the light cone determined by the optical metric $\bm{\frak{g}}$, measures of angle and distance are still understood as being made with respect to the background metric $\bm{g}$.
With this realization, one may go further than simple ray tracing to consider the behavior of congruences of light, i.e. beams in media \cite{Thompson2017prd}.
By analyzing such congruences, it has been shown that dielectric analog space-times -- the idea of mimicking the behavior of light propagation in a curved space-time with a corresponding dielectric residing in flat space-time -- are unfaithful in that although one may be able to mimic some idea of the ray trajectory of light, one cannot simultaneously replicate the behavior of a congruence \cite{Fathi2016prd}.

It should be emphasized that this result requires access to the background metric of the space-time in which the medium is embedded, which is of course the natural setting for the many ongoing experimental efforts in dielectric analog space-times.  
In a premetric setting, the optical metric is the only metric available and the analysis in that case is less clear, for while the optical metric of the vacuum would be identical to the optical metric of the background-free analog, the kinematics of congruences and the Raychaudhuri equation depend on covariant derivatives that in the vacuum are taken with respect to the background-metric compatible connection.
Presumably, one could by fiat impose a covariant derivative in a background-free analog that is compatible with the optical metric there and which might give agreement with the vacuum kinematics, at least in certain cases, e.g. where the initial manifold is vacuum and one chooses a trivial projection.
But if, for example, the initial manifold contains a nonvacuum contribution to the optical metric, such as from a refractive dust accretion disk in a Kerr background, then the distinction between background and optical metrics becomes immediately relevant in a determination of the kinematics -- a distinction that will \textit{not} carry through to the optical metric of a background-free analog. 
Such background-free constructions can therefore, in special cases, be consistent and distortion-free from a mathematical perspective, but are nearly always inconsistent from the perspective of real-world observers existing in background space-times making measurements with real instruments, and who may want to make other relevant (and in particular, timelike) measurements.

This approach to ray tracing in dielectric media in curved space-times may also be useful in astrophysical settings such as light propagating through an accretion disk that may possess some refractive properties.

\section{Summary and Conclusions} \label{Sec:Conclusions}

The goal of this paper was to study the covariant electrodynamics of general linear dielectric media within a curved space-time.
A $(3+1)$ space-time splitting was provided by a timelike vector field interpreted as a family of observers moving in the curved space-time.  This allowed the discussion to be formulated in terms of the familiar spacelike material parameters that would be measured by these observers.
Through the presence of the observers, the correspondence between $\bm{\chi}$ and the transverse $\bm{\varepsilon}$, $\bm{\mu}$, $\tensor[^h]{\bm{\gamma}}{}$ and $\tensor[^e]{\bm{\gamma}}{}$ was obtained through an explicit derivation that ensures true coordinate independence.

Of particular interest is the geometric optics limit and the derivation of the optical metric.
I followed a completely \textit{ab initio} derivation starting from the differential-forms expression of Maxwell's equations in linear media, and obtained an expression for the pseudo-Finslerian optical metrics in terms of the familiar material parameters.
I have shown that these pseudo-Finslerian optical metrics reduce to a pseudo-Riemannian optical metric for media obeying a curved space-time generalization of the Plebanski conditions, or for isotropic media.
For such media in curved space-times I have obtained an explicit formula for the optical metric, Eq.~(\ref{Eq:OpticalMetric}), and have shown that Gordon's optical metric is recovered exactly for isotropic media. 

The formulation pursued here is particularly suited to transformation optics \cite{Thompson2011jo1}, and it is anticipated that the structure studied here will provide a framework for expanding the transformation optics theory, for example to nonlinear, dispersive, and lossy media.
Furthermore, the ability to conduct ray tracing and study the kinematics of congruences through media in curved space-times could have application not only in transformation optics and dielectric analog space-times but also in astrophysical settings such as light (or radio wave) propagation through accretion disks around massive compact objects, or in dark matter, e.g. MACHO, surveys.

\subsection*{Acknowledgments}
I thank Mohsen Fathi, Jonathan Gratus, Paul Kinsler, and Martin McCall for useful discussions.
This work was supported in part by the Royal Society of New Zealand Marsden Fund Fast Start Grant No.\ UOO1219.

%


\end{document}